\documentclass{pnastwo}
\usepackage[numbers,sort&compress]{natbib}
\usepackage{amssymb,amsmath}

\usepackage{graphicx}
\usepackage{color} 

\definecolor{LinkColor}{rgb}{0,0,.5}
\definecolor{BrickRed}{rgb}{0.5,0,0}
\usepackage[colorlinks=true,linkcolor=BrickRed,citecolor=LinkColor,urlcolor=LinkColor]{hyperref}

\newcommand{\ham}{\mathcal{H}}
\newcommand{\half}{\frac12}

\newcommand{\ket}[1]{\left\vert{#1}\right\rangle}

\newcommand{\Id}{\leavevmode\hbox{\small1\normalsize\kern-.33em1}}
\newcommand{\carb}{$^{13}$C\ }
\renewcommand\text\mathrm

\begin{document}
\title{Atomic-scale nuclear spin imaging  using quantum-assisted sensors in diamond}
\author{A. Ajoy\affil{1}{Research Laboratory of Electronics and Department of Nuclear Science and Engineering, Massachusetts Institute of Technology,  Cambridge, Massachusetts 02139, USA},
U. Bissbort\affil{1}{Research Laboratory of Electronics and Department of Nuclear Science and Engineering,\\Massachusetts Institute of Technology,  Cambridge, Massachusetts 02139, USA}\affil{2}{Singapore University of Technology and Design, 138682 Singapore},
M.D. Lukin\affil{3}{Physics Department Harvard University}, R. Walsworth\affil{3}{}\affil{4}{Harvard-Smithsonian Center for Astrophysics and Center for Brain Science, Cambridge, Massachusetts 02138, USA}
\and P. Cappellaro\affil{1}{Research Laboratory of Electronics and Department of Nuclear Science and Engineering,\\Massachusetts Institute of Technology,  Cambridge, Massachusetts 02139, USA}}

\maketitle

\begin{article}
\begin{abstract}
Nuclear spin imaging at the atomic level is essential for the understanding of fundamental biological phenomena  and for applications such as drug discovery. 
The advent of novel nano-scale sensors has given hope of achieving the long-standing goal  of single-protein, high spatial-resolution structure determination in their natural environment and ambient conditions. In particular, quantum sensors based on the spin-dependent photoluminescence of Nitrogen Vacancy (NV) centers in diamond have recently been used to detect nanoscale ensembles of external nuclear spins. 
While NV sensitivity is approaching single-spin levels, extracting relevant information from a very complex structure is a further challenge, since it requires not only the ability to sense the magnetic field of an isolated nuclear spin,  but also to achieve atomic-scale spatial resolution. 
Here we propose a method that, by exploiting the coupling of the NV center to an intrinsic quantum memory associated with the Nitrogen nuclear spin, can reach a tenfold improvement in spatial resolution, down to atomic scales. The spatial resolution enhancement is achieved through coherent control   of the sensor spin, which  creates a dynamic frequency filter selecting only a few nuclear spins at a time. 
We propose and analyze  a protocol that would allow not only sensing individual spins in a complex biomolecule, but also unraveling couplings among them, thus elucidating local characteristics of the molecule structure.  
\end{abstract}
\maketitle

Proteins are the most important building blocks of life. 
The ability to obtain high-resolution protein structures is the keystone of drug discovery, since structure naturally reveals binding sites that can be targeted by drugs. 
Several methods exist for determining high resolution protein structure, primarily X-ray crystallography~\cite{Ramakrishnan00}, transmission electron microscopy~\cite{Auer00} and nuclear magnetic resonance (NMR)~\cite{Wuthrich86}. 
While each method has greatly contributed to our understanding of protein structure, none of them can sense individual molecules. Of particular interest would be imaging of small active sites  of a protein in its natural environment and conditions. 

Recently, novel quantum sensors associated with the nitrogen vacancy (NV) center in diamond~\cite{Taylor08} have shown the potential to provide few nT$/\sqrt{\textrm{Hz}}$ sensing  and nanoscale resolution at ambient conditions. NV-based magnetometers~\cite{Maze08,Balasubramanian08} can be transformative tools in quantum information, material science and bio-imaging~\cite{Cai13b,Hall12,Cooper14},  allowing for example high-resolution magnetic imaging of living cells~\cite{Lesage13}.
NV centers in bulk diamond have mapped the location of single $^{13}$C nuclear spins inside the diamond crystal~\cite{Taminiau12,Kolkowitz12a,Zhao12}, while shallow-implanted NVs have recently demonstrated the ability to sense a small number of nuclear spins in various organic materials~\cite{Mamin13,Staudacher13,Ohashi13,Mamin14}, as well as single and small ensembles of electronic spins outside the diamond sensor~\cite{Grinolds14,Ermakova13,Kaufmann13,Sushkov13u}.

\begin{figure}[b]
\centering
\includegraphics[width=0.45\textwidth]{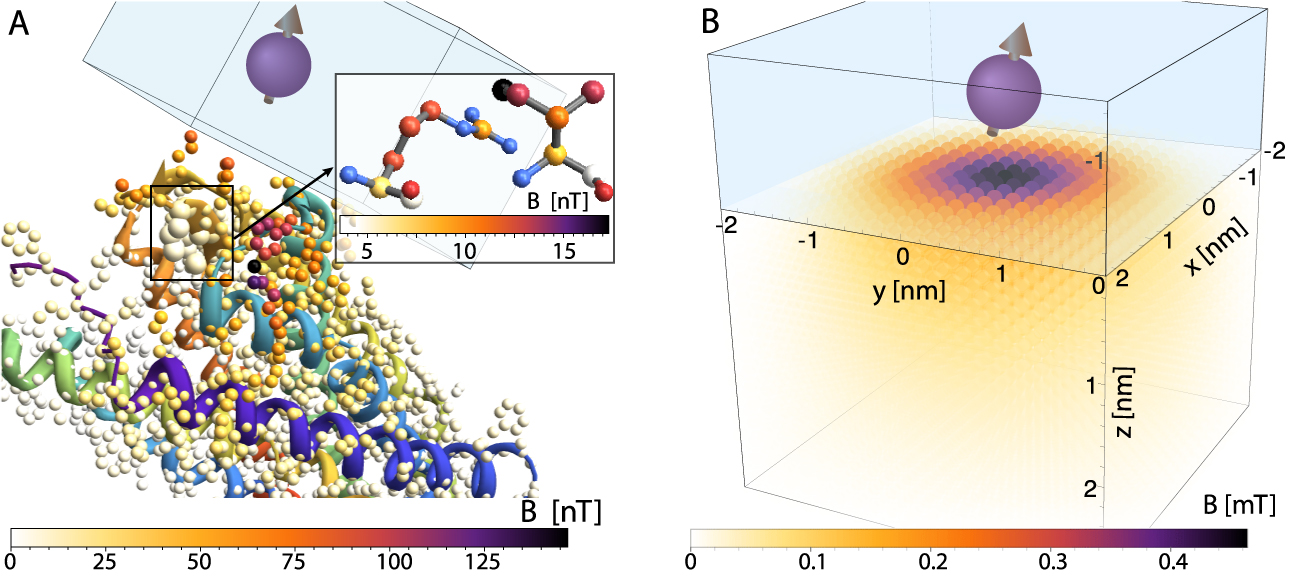}
\caption{(a) Nuclear spin imaging with a shallow NV center in diamond. A single NV spin (purple) at 1-2nm from the diamond surface can sense single nuclear spins in a molecule (the chemokine receptor CXCR4~\cite{Wu10}, ribbon diagram) anchored to the diamond. The magnetic field produced by individual \carb (spheres with color scale given by $B_\perp^j/\gamma_e$) is in the range of  nT, within reach of NV sensitivity. In the inset: the binding site of interest (atoms other than \carb are blue (O) and red (N)).
(b) A shallow NV center (2nm from the surface) creates a magnetic field gradient ($A(\vec r)/\gamma_n$) above the [111] surface of the diamond. Note  the azimuthal symmetry of the field, which causes degeneracy of the frequency shift at many spatial locations. \label{fig:NV-antenna}}
\end{figure}

Each nuclear spin in samples of interest creates a magnetic field that depends on its position relative to the NV center. For spins 2-5nm away from the NV spin, this magnetic field is within range of the sensitivity of the NV quantum probe~\cite{Taylor08} (see Fig.~\ref{fig:NV-antenna}.A), can be  made possible by the long coherence time achieved by careful material preparation~\cite{Balasubramanian09,Ohashi13,Staudacher12} and by using pulsed~\cite{Staudacher13} or continuous~\cite{Hirose12,Loretz13} 
dynamical decoupling (DD) sequences, which can extend the sensing time to hundreds of microseconds. 
The decoupling  cancels low-frequency dephasing noise, but by appropriately selecting the control timing to match the nuclear spin Larmor frequency, the NV spin can sense the magnetic field created by the nuclear spins,  even in the absence of a net nuclear polarization (spin noise). 
 For sparse nuclear spin ensembles, such as encountered in natural diamond,
this effect is enough to distinguish individual spins and thus to reconstruct their position~\cite{Taminiau12,Kolkowitz12a,Zhao12}.
For dense samples, 
the frequency resolution is inadequate to distinguish individual spins. In addition, the intrinsic NMR linewidth in a dense molecule might exceed the NV sensing linewidth.

In the following we describe a method to improve the frequency, hence spatial, resolution of NV sensing. We evaluate the method performance with simulations on typical bio-molecules and discuss the experimental resources needed to achieve atomic-scale reconstruction of nuclear spin positions.

\subsection{Principles}
We assume that the biomolecule to be probed is attached to the surface of a diamond at the location of a single NV center that is within 3 nm of the diamond surface (Fig.~\ref{fig:NV-antenna}).  To achieve enhanced spatial resolution for NV sensing of the nuclear spins within the biomolecule, we use three critical ingredients: (i) an effective strong magnetic field gradient across the biomolecule due, e.g., to the rapid spatial variation of the magnetic dipolar field from the NV electronic spin; (ii) Hamiltonian engineering of the NV and multi-nuclear spin system~\cite{Ajoy13l} -- which creates a very sharp dynamical frequency filter for NV sensing and controls the effect of NV spin polarization leakage through the coupled nuclear spin network; and (iii) a quantum memory -- the \textit{nuclear} $^{15}$N spin associated with the NV center -- which enables long acquisition times and hence an even sharper frequency filter.

The sensing protocol works by following the leakage of polarization from the NV center (which is optically polarized) toward on-resonance nuclear spins. The resonance condition is in principle different for each nuclear spin, as it is given by its position-dependent coupling to the NV center. However, simply implementing cross-polarization does not yield enough frequency resolution for single-spin sensing.
By alternating periods of cross-polarization with evolution under a magnetic field gradient --while the NV spin state is preserved by mapping onto the quantum memory-- the interaction Hamiltonian is modulated and only couplings of the NV to resonant nuclear spins are retained. Evolution under the gradient thus acts as a sharp frequency filter, thereby increasing the spatial selectivity. The gradient evolution time can be long, limited only by the quantum memory (the \textit{nuclear} $^{15}$N spin associated with the NV center) coherence time, which at a few milliseconds translates to a high frequency resolution, $\delta A\sim100$Hz.

Using the NV polarization leakage as a signature of the nuclear spin position provides two advantages over measurement of  NV frequency shifts due to specific nuclear spins. First, it is easy to embed nuclear decoupling sequences within the sensing protocol; nuclear decoupling is essential to narrow the NMR spectrum intrinsic linewidth and thus detect signal arising from individual nuclear spins. Second, once the polarization is transferred to a nuclear spin, this spin itself becomes a sensitive probe of its local environment. 
In the following we propose a three-step protocol for nuclear spin imaging. In a first step, we observe polarization transfer from the NV center to one (or a few) nuclear spin(s), providing partial information about their positions. The nuclear spins can then evolve freely, and polarization diffuses in the nuclear spin network. The polarization is then transferred back to the NV center, revealing the nuclear spin connectivity.
 We can thus acquire multi-dimensional spectra that not only decrease frequency crowding, but can also resolve degeneracies related to the intrinsic symmetry of the dipolar coupling.  

Next we describe in detail the proposed method and its application to biomolecule structure reconstruction.

\section{Results}
\subsection{Nuclear Spin Detection via NV centers in diamond}
The electronic spin $S=1$ associated with the negatively charged Nitrogen Vacancy center in diamond is a sensitive probe of magnetic fields at the nano-scale~\cite{Taylor08}. The good sensitivity of NV spins is due to the long coherence time achieved under dynamical decoupling.
Pulsed decoupling sequences -- a train of $\pi$-pulses such as CPMG~\cite{Meiboom58} -- not only cancel low-frequency dephasing noise, but also act as sharp band-pass filters at selected frequencies set by the pulse timing~\cite{Biercuk11,Ajoy11}.

Nuclear spins on the surface of the diamond couple to the NV electronic spin via the magnetic dipolar interaction $\vec\mu$,
\begin{equation}
\vec\mu(\vec r) = \frac{\mu_0}{4\pi}\frac{\hbar}{2\pi}\frac{\gamma_e\gamma_n}{r^5}\left[(3r_z^2 -r^2)\hat z+ 3 r_z(r_x \hat x + r_y\hat y)\right],
\label{eq:dipolar}
\end{equation}
where the NV is at the origin and we aligned the coordinate system with the NV symmetry axis. Here $\gamma_{e,n}$ are the gyromagnetic ratios of the electronic and nuclear spins respectively.
For shallow NVs, this interaction   exceeds the nuclear-nuclear dipolar coupling and is strong  enough to allow coherent coupling to individual nuclear spins, via the Hamiltonian:
\begin{equation}
\ham^j_{dip} = S_z\vec\mu(\vec r_j)\cdot\vec I^j= S_z[A^jI^j_z+B^j_\perp(\cos\phi^jI_x^j+\sin\phi^jI_y^j)],
\label{eq:HamDip}
\end{equation}
 where $S_\alpha$  and $I^j_\alpha$ are the NV electronic spin and external nuclear spin operators, respectively.
Here  $A^j=\mu_{z}(\vec r^j)$ and $B^j_\perp=\sqrt{\mu_{x}^2+\mu_{y}^2}$ are the longitudinal and transverse components of the dipolar coupling at the location of the j$^{\text{th}}$ spin.

 The normalized signal due to the net phase shift imparted to the NV spin after a train of $2n$ $\pi$-pulses spaced by a time $\tau$ is given by
$S=\half\left(1+\prod_j \mathcal S^j\right)$, where the pseudo-spin signal from the j$^{\text{th}}$ nuclear spin is
\begin{equation}
 \mathcal S^j=\!1\!-\!2 \vec\omega^j_0\!\times\!\vec\omega^j_1\sin^2\!\left(\!\frac{\Omega^j_0\tau}{4}\!\right)\sin^2\!\left(\!\frac{\Omega^j_1\tau}{4}\!\right)\frac{\sin(n\alpha^j)^2}{\cos(\alpha^j/2)^2},
 \label{eq:SCP}
\end{equation}
\[  \cos(\alpha^j)\!=\!\cos\!\left(\!\frac{\Omega^j_0\tau}{2}\!\right)\!\cos\!\left(\!\frac{\Omega^j_1\tau}{2}\!\right)\!-\!\vec\omega^j_0\cdot\vec\omega^j_1\sin\!\left(\!\frac{\Omega^j_0\tau}{2}\!\right)\!\sin\!\left(\!\frac{\Omega^j_1\tau}{2}\!\right)\!.
\]

Here the vectors $\vec \Omega^j_i=\Omega^j_i\vec{\omega}^j_i$ represent the nuclear spin Hamiltonian in the two subspaces of the NV electronic spin.
In general we have $\vec\Omega^j_0=\omega^j_L\hat z$, where $\omega_L=\gamma_nB$ is the nuclear spin Larmor frequency. Then the signal shows ``dips'' around $\tau_k=(2k+1)\pi/\omega_L$ marking the presence of nuclear spins. However, the dip times are specific for each nuclear spin because of the differences in dipolar coupling, $\vec \Omega^j_1=(\omega_L+A^j)\hat z+B_\perp^j[\cos\varphi^j\hat x+\sin\varphi^j\hat y]$. 
The dip minimum is achieved when $\alpha^j=\pi$; for $\vec\omega_0^j\cdot\vec\omega_1^j\approx1$ (as is the case for large enough magnetic field); this is obtained at $\tau_k^j=2\pi(2k+1)/(\Omega_0^j+\Omega_1^j)\approx(2k+1)\pi/(\omega_L+A^j/2)$. At long times, the signal dips arising from different nuclear spins become discernible,
as one can separate contributions from spins having dipolar couplings $A^j-A^i=\delta A\gtrsim\omega_L/[(2k+1)n]$. The linewidth of this sensing scheme is thus limited by the coherence time,
$\delta A\sim 4\pi/(T_{2})\approx(2-50)2\pi$~kHz.  While the signal contrast at the dips,  
$S^j\sim \cos\left(\frac{2B^jn}{\omega_L+A^j/2}\right),$ would be enough to measure protons 2-5nm from the NV center, the frequency resolution is not enough to distinguish individual spins separated by $\sim0.1$nm in a dense molecule. In addition, the intrinsic NMR linewidth of dense samples (tens of kHz) can exceed the NV sensing linewidth and it is challenging to embed homonuclear decoupling in the NV decoupling-based sensing scheme (SI Appendix). 

An alternative approach to nuclear spin sensing is to observe the polarization leakage from the NV center to the nuclear spins
~\cite{London13,Belthangady13}. 
We first rotate the NV spin to the transverse axis and then apply a continuous driving along that same axis (spin-locking). The driving decouples the NV from noise and the nuclear spin bath. Setting however the NV  Rabi frequency $\Omega$ close to the target nuclear spin energy, $\Omega\approx\pm\omega_L$,  polarization is transferred to the nuclear spins~\cite{Hartmann62,Henstra88}, resulting in a signal dip:
\begin{equation}
S^j=1-\frac{B_\perp^{j2}\sin^2\left(\frac{t}{2}\sqrt{B_\perp^{j2}+[\Omega\mp(A^j+\omega_L)]^2}\right)}{2\left[B_\perp^{j2}+[\Omega\mp(A^j+\omega_L)]^2\right]}.
\label{eq:HH}
\end{equation}
The energy matching condition depends on the nuclear spin dipolar coupling $A^j$, but the frequency resolution is limited by the hyperfine transverse component, $\delta A\!\sim\!B^j$. Even if  homonuclear decoupling can be embedded within cross-polarization, it is not possible to distinguish spins in a dense sample. Also, cross polarization is often plagued by power fluctuation in the driving field, which leads to imperfect energy matching and broadens the effective frequency resolution.

\subsection{Quantum-enhanced spatial resolution}
To increase the frequency --and thus spatial-- resolution, we exploit the presence of an ancillary qubit associated with the Nitrogen nuclear spin ($^{15}$N with $I=\half$). This long-lived ancillary spin can store information about the state of the NV electronic spin, while the external nuclear spins  evolve under the action of a magnetic field gradient, which creates a further frequency filter, thereby increasing the spatial selectivity.
\begin{figure}[b]\centering
\includegraphics[width=0.45\textwidth]{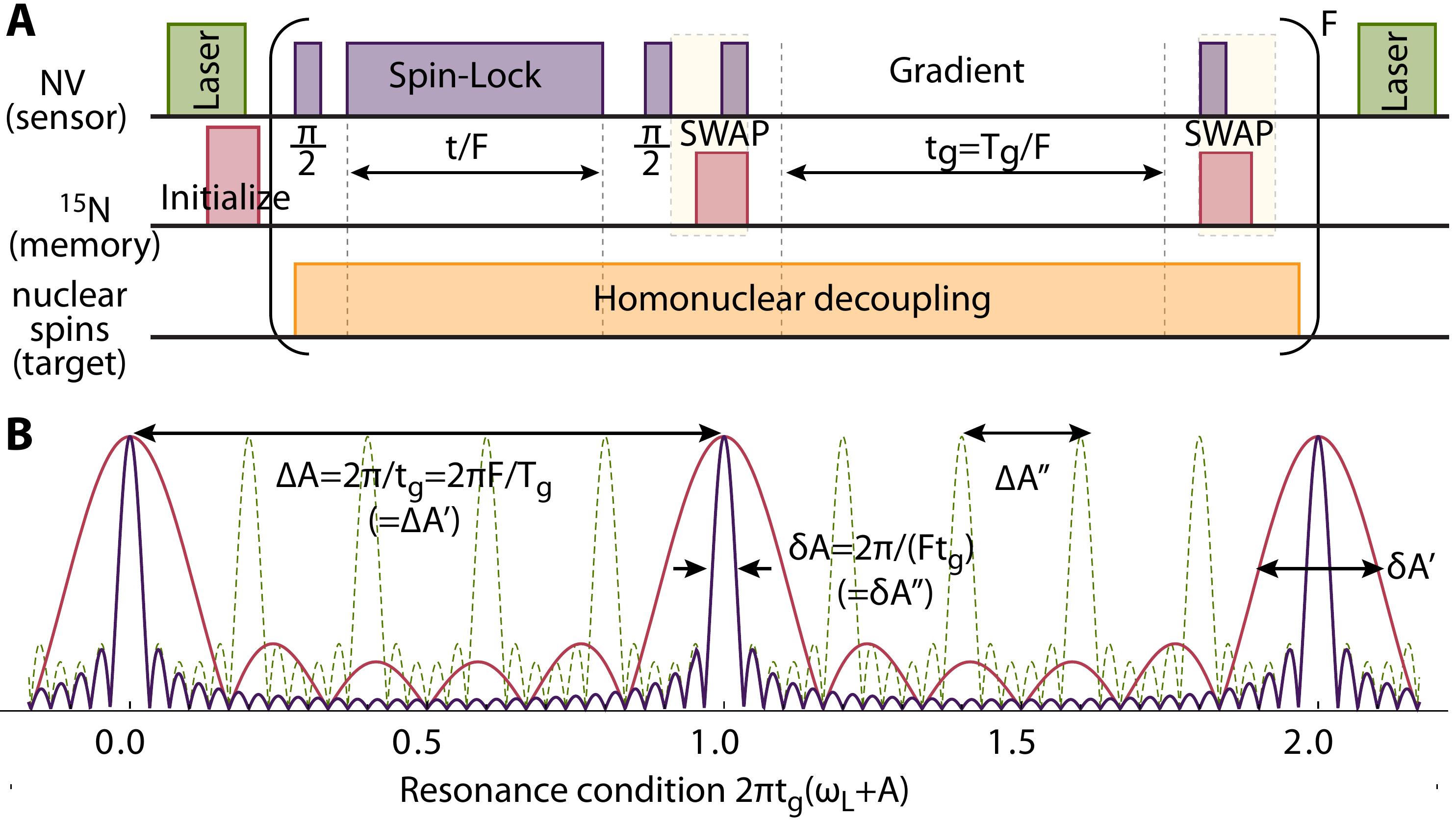}
\caption{A: Control scheme for filtered cross-polarization nuclear spin sensing. Polarization is transferred  from the NV electronic spin to the target nuclear spins in the bio-molecule for a time {\small$t/F$}. To increase spatial resolution, the nuclear spins are let to evolve under a magnetic field gradient for a time {\small$t_g=T_g/F$}, while the NV state is stored in the nuclear {\small$^{15}$}N spin. The scheme is repeated {\small$F$} times to create a sharp filter. For reverse-sensing the two {\small$\pi/2$} pulses are omitted.
 B: Filter created by the control scheme as a function of the normalized time {\small$2\pi t_g(\omega_L+A)$}. The filter peaks are sharper for higher {\small$F$} (dark line, {\small$F=30$}, red line, {\small$F'\!=\!6$}), yielding a smaller linewidth {\small$\delta A$}. While it is possible to achieve sharper peaks at longer times (dashed line, {\small$F''\!=\!6$, $t_g''\!=\!5t_g$}), this results in a smaller bandwidth {\small$\Delta A''$}. \label{fig:Filter}}
\end{figure}
The gradient evolution time can be as long as a few milliseconds,  limited only by the \textit{nuclear} $^{15}$N spin coherence time, yielding a sharp frequency filter.
The gradient field could be provided by a strong magnetic tip~\cite{Grinolds14,Degen09}, micro-fabricated coils~\cite{Arai13} or dark spins~\cite{Schaffry11} on the diamond surface. Even more simply, the NV itself can create the gradient, when set to the $\ket1$ state (see Fig.~\ref{fig:NV-antenna}.B).
In the following, we will assume this strategy,  as this does not require any additional experimental resource.

The method can proceed as follows. After a period of evolution under the spin-lock Hamiltonian, the NV state is mapped onto the nuclear ancillary spin --initially in the $\ket{1}$ state -- by a SWAP gate~\cite{Nielsen00b} (see Fig.~\ref{fig:Filter}).
 High fidelity SWAP operations  can be constructed by successive RF and microwave irradiation~\cite{Jiang09}, by free evolution under a misaligned field or via the use of decoherence protected gates~\cite{Cappellaro09,van12}. The electronic spin is thus left in the  $\ket{1}$ state, thereby creating a magnetic field gradient. Inverting these operations after a time $t_g$ effectively implements an evolution of the external nuclear spins under the Hamiltonian
\begin{equation}
H_G=\Id_{NV}\otimes\textstyle\sum_j H_G^j,\quad H_G^j\approx (\omega_L+A^j )I^j_z.
\label{eq:gradient}
\end{equation}
Alternating evolution under the gradient and the spin-lock Hamiltonian (in the dressed-state basis),
\begin{equation}
H_{\text{SL}}=\Omega S_z+\textstyle\sum_j \left[(\omega_L+A^j)I^j_z +  B^j_\perp (e^{i\phi_j}S_+I_-^j+\text{h.c.})\right],
\label{eq:spinlock}
\end{equation}
where $I_\pm=I_x\pm i I_y$, we obtain the effective evolution:
\begin{equation}
\left[e^{-i H_{G}t_g}e^{-iH_{\text{SL}}t/F}\right]^F \!\equiv\! \exp(-iH_Gt_gF)\exp(-it\overline{H}_{\text{F}})
\label{eq:evolution}
\end{equation}
Here we define the filtered Hamiltonian  $\overline{H}_{\text{F}}\!=\!\sum_j \overline{H}_{\text{F}}^j$, with
\begin{equation}
\overline{H}_{\text{F}}^j=\Omega S_z+(\omega_L+A^j)I^j_z +B^j_\perp (\mathcal{G}^je^{i\phi_j}  S_+I_-^j+\text{h.c.}),
\label{eq:avHam}
\end{equation}
\begin{equation}
{\cal G}^{j}(t_g)\!=\! \frac1F\sum_{k=0}^{F-1}e^{ikt_g(\omega_L+A^j)},\quad |{\cal G}^{j}|\!=\!\frac{\left|\sin \left(\frac{F t_g(A^j+\omega )}{2} \right)\right|}{F \left|\sin \left(\frac{t_g(A^j+\omega )}{2} \right)\!\right|}
\label{eq:filter}
\end{equation}
(see Methods). This control scheme creates a time-domain Bragg grating~\cite{Ajoy11} around the frequencies $\omega_f=2n\pi/t_g$, with very sharp peaks for large $F$ (see Fig.~\ref{fig:Filter}.B). Then, only for on-resonance nuclear spins the dipolar coupling transverse component, $B_\perp^j$, is retained and polarization is transferred, yielding $S^j<1$ in Eq.~(\ref{eq:HH}).
The filter linewidth, $\delta A =  2\pi/(F t_g)$, is determined by the gradient-evolution time, which can be made very long since it is only limited by the ancillary nuclear spin dephasing time, $Ft_g\lesssim T_{2n}^*\sim 8-10$ms. The bandwidth $\Delta A$ over which it is possible to distinguish different spins is instead set by the condition $(A\pm\Delta A+\omega_L)t_g\lessgtr 2\pi(n\pm1)$.  The scheme can thus achieve a frequency resolution of about $\delta A\sim2\pi100$Hz over a bandwidth $\Delta A = 2\pi/t_g\sim F\times2\pi100$Hz, which can be 10-50 times larger than $\delta A$. The bandwidth can be further improved by over five times using a novel control strategy that effectively creates  anti-aliasing filters~\cite{Ajoy13l} (SI Appendix).
Since it is possible to embed homonuclear decoupling~\cite{Waugh68} during both the spin-lock and the gradient evolution, the linewidth is not limited by the spin-spin coupling (see Methods). We note that a similar filtering scheme could be embedded with the DD-based sensing strategy (Eq.~\ref{eq:SCP}), increasing its frequency resolution while also allowing homonuclear decoupling during the gradient evolution.

The ancillary $^{15}$N nuclear spin, by acting as a memory, thus achieves two goals:  it increases the coherence time and it enables using the NV center as a source of magnetic field gradient. The resulting dynamic Bragg grating reaches high spatial resolution (sub-angstrom for protons and a NV 2-3nm below the diamond surface). As shown in Fig.~\ref{fig:simulation}, the frequency resolution is much better than for DD-based sensing protocols.

A further advantage of the proposed filtered cross-polarization scheme is that it can be used not only to sense individual nuclear spins, but also to polarize them. The nuclear spins, as explained below, then become local probes of their environment, providing essential structural information.

\subsection{Protocol for nuclear spin imaging}
\label{sec:protocol}
\begin{figure}[b]\centering
\includegraphics[width=0.3\textwidth]{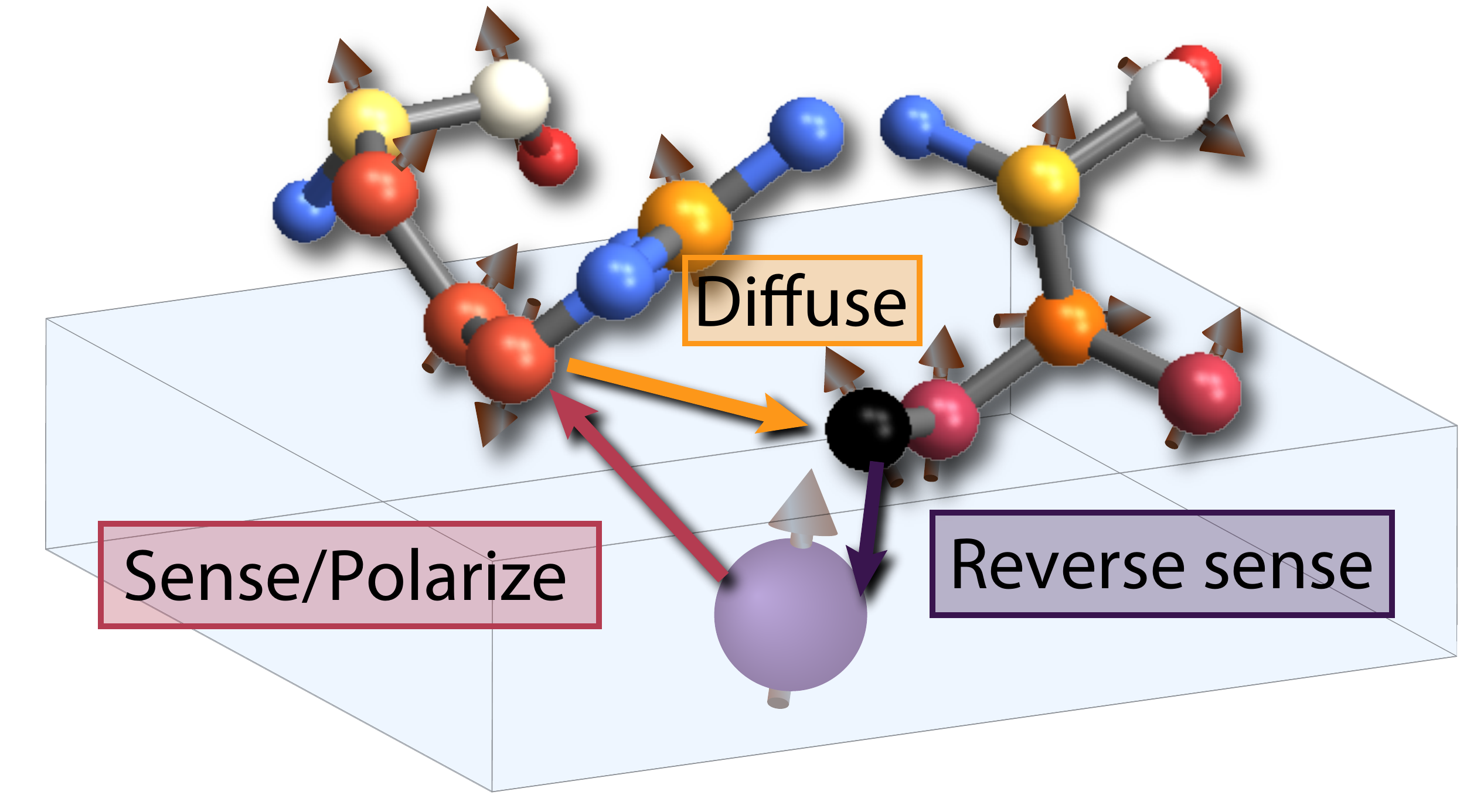}
\caption{Protocol for quantum-enhanced nuclear spin imaging. First the NV measures a 1D NMR spectrum  ({sense}). During this step, polarization is selectively transferred from the NV to a particular nuclear spin in the protein ({polarize}). Polarization is then allowed to spread ({diffusion}), driven by the nuclear spin-spin dipolar coupling. The polarization now localized  on a different  nuclear spin is transferred back to the NV spin and measured optically ({reverse-sense}). 
\label{fig:expt-protocol}}
\end{figure}

A  protocol for nuclear spin imaging involving three steps is illustrated in Fig.~\ref{fig:expt-protocol}. First, we acquire a 1D NMR spectrum  using  the filtered  cross-polarization method described above (``\textit{sense/polarize}'' step), sweeping the filter time $t_g$ and the driving frequency $\Omega$. For each time point, only one (or a few) nuclear spin(s) becomes polarized and thus contributes to the signal.
In the second step, the nuclear spins are left to evolve freely for a time $t_d$ (``\textit{diffuse}'') prior to transferring polarization back to the NV electronic spin for detection in the third step (``\textit{reverse sense}''). During the diffusion time $t_d$, nuclear polarization migrates to neighboring spins under the action of the homonuclear dipolar Hamiltonian~\cite{Khutsishvili66}. To sense the new location of the polarization the NV is again driven at the nuclear spin frequency (alternating with the gradient field for enhanced spatial resolution). By omitting the initial rotation to the transverse plane, the NV is now only sensitive to polarized nuclear spins. 
 For a fixed diffusion time $t_d$ one thus obtain a 2D, correlated spectrum, similar to NMR 2D spectroscopy~\cite{Aue76}. The spectrum (Figure~\ref{fig:simulation}) encodes information about the couplings between nuclear spins and thus constrains their relative positions. 
 Note that we could further obtain 3D spectra by sweeping the diffusion time $t_d$ (SI Appendix).  The spectra yield an over-constrained system for the spin positions, as for $N$ nuclear spins we can obtain up to $2N+N(N-1)/2$ equations (SI Appendix). By using the polarized nuclear spins as probes of their own environment, this protocol can thus break the symmetry of the dipolar coupling and overcome the phase problem that plagues other experimental techniques. 

Indeed, the 1D NMR spectrum acquired in the first step provides information on the dipolar couplings to the NV spin. Using a peak-picking procedure, we directly read out the longitudinal dipolar couplings from the frequency position $\omega_p$ of the dips, $A\!=\!2(\omega_p\!-\!\omega_L)$. The dip height determines the transverse coupling, 
as $S^p\!=\!1\!-\!\half\sin^2\!(B_\perp t/2)$ (Eq.~\ref{eq:HH}). Given a pair of parameters $A_p,B_p$ for each dip, the spin position is found by inverting Eq.~(\ref{eq:dipolar}). This yields a pair of solutions for $r_z$ and $r_\perp=\sqrt{r_x^2+r_y^2}$, with only one usually consistent with the NV depth. 
The nuclear spin positions lie in a region defined by the unknown angle $\phi=\arctan{(r_y/r_x)}$ (analogous to the phase problem in x-ray crystallography.) While the NV depth often further constrains this region, the information obtained from the 2D spectrum correlations is essential to unambiguously determine the position. 
In addition, 2D correlations help in distinguishing spins that might have similar couplings to the NV (because of the phase symmetry) and thus create overlapping signals in the 1D spectrum, but give rise to distinct peaks in the 2D spectra since they couple to different spins.

For an isolated nuclear spin pair, the signal intensity at the diagonal peaks is $S_{ii}\!=\!\half[1+\sin(B_\perp^jt/2)^4\cos(D_{ij}t_d/2)^2]$, where $D_{ij}$ is the dipolar coupling strength between the spins. The cross peaks (yielding a cleaner signal for non-isolated pairs) are instead  $S_{ij}\!=\!\half[1+\sin(B_\perp^jt/2)^2\sin(B_\perp^it/2)^2\sin(D_{ij}t_d/2)^2]$. 
For each spin pair,  knowledge of their mutual dipolar coupling $D_{ij}$ determines the phase difference $\phi_i-\phi_j$, thus all relative phases can be determined (up to a global phase) if enough dipolar couplings can be measured. 
In general, a 2D spectrum might not be sufficient to fully determine the position of a complex structure since couplings between distant nuclear spins might not be visible and because of uncertainties in the estimated parameters. The information can however be supplemented by 3D spectra (varying the diffusion time) and by including prior information, such as the length of chemical bonds, into the reconstruction algorithm.

\section{Discussion}
The proposed method would enable NV centers in diamond to sense individual nuclear spins and their mutual couplings with atomic-scale resolution and thereby determine the atomic structure of pockets and local sites close to the surface of large bio-molecules. Experiments can be performed at ambient conditions, without the need to crystallize the molecules, and on single molecules, avoiding the need to synthesize large ensembles.  Site-selective isotopic labeling~\cite{Kainosho06} would allow focusing on a small number of $^{13}$C or Nitrogen spins, or even protons in molecules in a deuterated solution. 
The molecules could be functionalized and attached to the diamond surface, for example by methods such as the EDC/NHS reaction, which forms a chemical bond  stable over a few weeks~\cite{Sushkov13u}. 
Deterministic binding near an NV center may be obtained using co-localization techniques developed in~\cite{Sushkov13u}. 

Measuring the structure of local surface sites in biomolecules, relevant to many biological functions and drug discovery, is within reach of current NV sensitivity, although the mapping of whole bio-molecules is more challenging and requires further technical advances in quantum metrology~\cite{Kessler13x} or nano-diamonds with enhanced material properties~\cite{Trusheim13}. Our method could complement information acquired via existing techniques, in particular x-ray crystallography and NMR. While x-ray diffraction is able to image whole proteins with Angstrom resolution, the technique requires the ability to produce high quality single crystals of a few hundred of microns, thus restricting the number of proteins that can be studied (e.g., membrane proteins cannot be crystallized); in contrast, NV-based sensing can image single proteins in their natural state. In addition, the imaging protocol we propose can help solve reconstruction issues associated with the phase problem in x-ray diffraction and overcrowded spectra in NMR. While the NV-based scheme has similarities to NMR methods such as NOESY~\cite{Wuthrich86}, which explore through-space spin-spin correlations, the ability of our NV technique to polarize only one spin at a time allows acquiring more information even in possibly crowded spectra. 

We can estimate the resources needed for such tasks by considering the frequency resolution and signal-to-noise ratio (SNR) requirements to distinguish one or few nuclear spins in a dense molecule. For a nuclear spin on-resonance with the NV Rabi frequency, the signal in Eq.~(\ref{eq:HH}) simplifies to $S^j=1-\half\sin^2(B_\perp t/2)$, where the time $t$ is limited by the coherence time in the rotating frame ($T_{1\rho}\approx 2$ms~\cite{Loretz13}). 
The SNR in $M$ measurements is then $SNR\sim \sqrt{M/2}CT_\rho$, where $C\approx0.01-0.3$~\cite{Taylor08,Jiang09} captures the finite contrast and photon collection efficiency.  
In order to distinguish spins with dipolar couplings differing by $\delta A$, the total gradient time should be $T_g=Ft_g=2\pi/\delta A$ in $F$ steps. Thus the total time required to measure one spin is $T_s=M(T_g+T_\rho+Ft_a+t_{\text{ro}})$, where $t_{\text{ro}}$ is the read-out time and $t_a$ is the time needed for the ancillary spin protocol (polarization of the NV, $\sim$500ns, NV $\pi$-pulse, $20$ns and SWAP gate, $\sim 4\mu$s).
For the 2D protocol, assuming division of  the spectral range in $b$ frequency steps and letting the nuclear spin diffuse for a time $t_d$, the total measurement time becomes $T_{2D}=b(T_{1D}+t_d)=b^2T_s+bt_d$. 
\begin{figure}[th]
\centering
\includegraphics[width=2.9in]{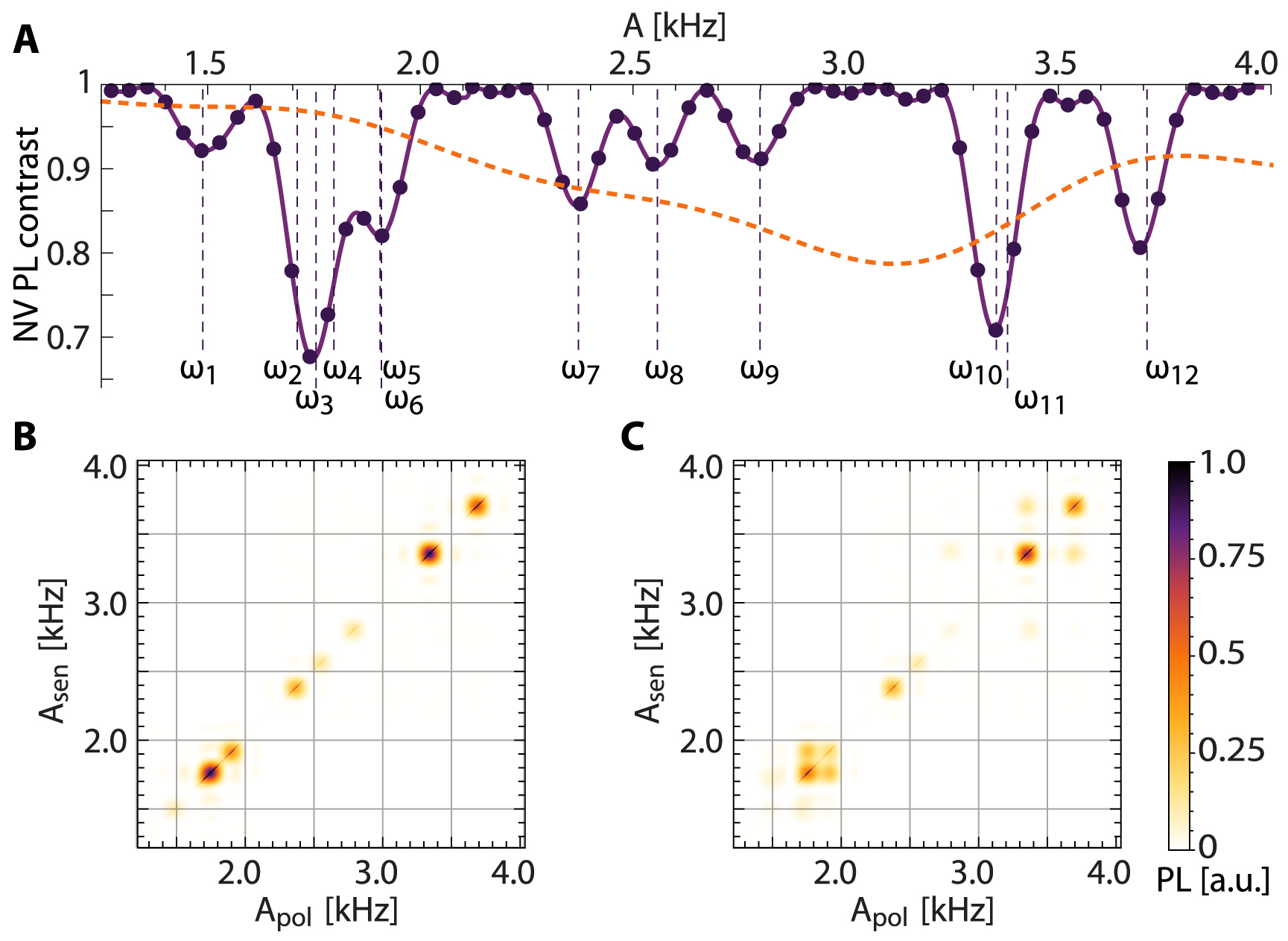}
\caption{Simulated 1D and 2D NMR spectra for the binding site in CXCR4, obtained with the NV-based filtered sensing protocol. A: Simulated normalized spin-dependent NV photoluminescence (PL)  after the filtered cross-polarization sequence.  The x-axis shows the on-resonance dipolar frequency $A$ at each measurement point, while the driving frequency is {\small$\Omega\!=\!A/2/\sqrt3+\omega_L$} (with Larmor frequency $\omega_L=2$MHz) and the gradient time {\small$t_g\!=\!2\pi/(A/\sqrt3+\omega_L$)}. A homonuclear decoupling sequence was applied to the \carb spins to narrow their intrinsic linewidth. The PL shows dips when $A$ and the inverse gradient time (simultaneously swept) match the longitudinal dipolar coupling {\small$A^j$} of one nuclear {\small$^{13}$}C spin in the protein.
We simulated the 12 {\small$^{13}$}C spins in the ARG and ILE active site and an NV center 1.75nm below the diamond surface, with the [111] axis aligned around the vertical direction. The gradient time was varied from 201.68 to 201.84$\mu$s (for a maximum total gradient time of {\small$T_g\!\approx\!6$}ms in {\small$F\!=\!30$} steps). The total polarization time was {\small$t=720\mu$}s. The dip height is a measure of the transverse dipolar coupling {\small$B^j_\perp$}. The spectrum shows 8 dips, with the height indicating overlapping contributions from almost equivalent spins. The dashed line is the spectrum one would obtain with a DD-based  protocol distorted by the spin-spin couplings (SI Appendix). B: Simulated 2D NMR spectra from the proposed protocol. After polarization (obtained with the same parameters as for the 1D spectrum) the nuclear spins  evolve freely for {\small$t_d\!=\!300\mu$}s, allowing polarization diffusion. The polarization is then mapped back to the NV center and measured via its spin-dependent PL. The left plot is the 2D NMR spectrum for no diffusion {\small$t_d\!=\!0$}, while the right plot shows the spreading of polarization in the nuclear spin network, as indicated from off-diagonal non-zero terms. \label{fig:simulation}}
\end{figure}

Consider for example the chemokine receptor CXCR4~\cite{Wu10}.  Chemokine receptors are G protein-coupled receptors  found predominantly on the surface of leukocytes and are critical regulators of cell migration for immune surveillance, inflammation, and development. In particular, the CXCR4 receptor (Fig.~\ref{fig:NV-antenna}.A) is implicated in cancer metastasis and HIV-1 infection. CXCR4 ligand-binding cavities are actively studied as mutations are reported to decrease HIV-1 infectivity. CXCR4 was recently crystallized and its structure determined by x-ray diffraction~\cite{Wu10} (we use these data in our simulations). It would be desirable to acquire more information about  conformational changes in these receptors,  resulting in signal transduction and HIV-1 entry process, as well as to obtain information about similar membrane proteins that cannot be crystallized.

In order to estimate the resources required to determine the structure of CXCR4 binding sites, we simulated the dynamics of the NV electronic spin under the action of the proposed control sequences and the protein's nuclear spins. In the simulations we assumed that only the amino-acids ARG and ILE are $^{13}$C-labeled, thus we can focus just on the 12 $^{13}$C nuclear spins of the 183 and 185 residues, as coupling to other spins (e.g., protons) is off-resonance or too weak. We further assumed application of both heteronuclear and homonuclear decoupling sequences  to narrow the $^{13}$C linewidth.
For an NV 1.75nm below  the diamond surface, the minimum dipolar coupling of interest is $B_\perp=250$Hz. 
Thus we need $M\approx1000$ 
acquisitions to see such a spin, assuming a collection efficiency of $C\approx0.2$ which can be obtained by 100 repeated readouts of the electronic spin, using the ancillary nuclear spin~\cite{Jiang09,Neumann10b,Ajoy12g}. 
Choosing $b=15$ bins to scan the frequency bandwidth of interest, $W=2.5$kHz, 
the gradient time for the target linewidth is $T_g=6$ms$=1/(166\text{Hz})$. The experimental time to acquire a 1D spectrum  is about $T_{1D}\approx2$ minutes, while mapping the 2D correlations for the 10-12 frequencies of interest  requires about $T_{2D}\approx 30$minutes. 
In Fig.~\ref{fig:simulation} we show simulated 1D and 2D spectra for the CXR4 receptor active site; to illustrate the potential of our filtered scheme we increased the number of frequency bins to $b=64$. The figure shows the high frequency resolution the method can achieve as well as the additional information provided by the local diffusion of polarization in the nuclear spin network, resulting in position estimates for each resolvable spins with good precision. Assuming an uncertainty of $\Delta A=\Delta B=300$Hz for the dipolar couplings and an error $\sqrt{ [\Delta D_{ij}=0.5 D_{ij} ]^2+ [50\mbox{Hz}]^2 }$, the typical volume uncertainty ranges from $1.2$\AA$^3$ to 10 \AA$^3$ for the various spins we considered (see SI Appendix for details), although it increases sharply with the distance from the NV. 

We note that these estimates do not include atomic position fluctuations due for example to thermal motion. The assumption that the atoms are quasi-static during the duration of the experiment is nonetheless reasonable, especially for nano/micro-crystalline structures.
In general, obtaining a precise estimate of atom fluctuations is difficult, since it depends on several factors, including molecule topology and torsion, binding mechanism to the diamond surface and surrounding molecular environment (liquid hydration layer  or ice). However, we can obtain estimates for crystal structures from the Debye-Waller coefficients tabulated in the PDB -- at 77K, the disorder in atom positions for most crystals is under 1.5\AA, comparable to the sensing resolution of our protocol.

\section{Conclusion}
We proposed a practical method for atomic-scale nuclear spin imaging in bio-molecules using NV centers in diamond.
Recent developments in materials fabrication~\cite{Ohno12, Balasubramanian09}, ion implantation~\cite{Osterkamp13,Naydenov10} and coherent control techniques~\cite{Naydenov11,DeLange10} have brought diamond magnetometers close to the threshold of single nuclear spin sensitivity. These quantum sensors have the potential to be an important tool in proteomics, as they overcome some of the challenges plaguing other experimental techniques,  such as x-ray diffraction and conventional NMR. Most prominently, they would not require crystallization of the sample, a challenge for many classes of bio-molecules such as membrane proteins; nor large sample sizes.

Our novel strategy combines coherent control of the NV sensor with an intrinsic quantum memory to enhance the sensor spectral resolution.
This control strategy not only creates a sharp dynamic filter by alternating periods of a spin-lock Hamiltonian with evolution under a gradient field, but  provides  other advantages. The sequence is compatible with homonuclear decoupling, thus allowing sensing beyond the natural bio-molecule NMR linewidth. In addition,  our technique allows mapping the couplings among the spins themselves, using them as local probes of their environment. The resulting multi-dimensional NMR spectra highlight spatial correlations in the sample, lift spectral overlaps due to symmetries and aid the structure reconstruction algorithms.
This would allow resolving  the contributions in the NV signal arising from different nuclear spins in a dense sample  and to use the acquired information to determine the nuclear spin positions. Reconstructing  a protein local 3D structure in its natural conditions would allow   researchers to work backwards and design compounds that interact with specific sites.
 
By combining the strength of NMR-inspired control techniques with the quantum properties of NV center spins, the proposed strategy for magnetic resonance detection at the nano-scale promises to make diamond-based quantum sensors  an invaluable technology for bio-imaging.

\begin{materials}
\subsection{The Qubit Sensor}
The Nitrogen-Vacancy color center in diamond is a localized defect in the diamond lattice comprising a Nitrogen substitutional atom adjacent to a vacancy. The negatively charged NV that we consider here forms a spin 1 in its ground state, with a zero-field splitting of {\small$\Delta\approx2.87$}GHz, which determines the system's eigenstates. An additional magnetic field along the NV axis further splits the degeneracy between the $\ket{\pm}$ states and only the transition between the $0$ and $-1$ states is addressed with on-resonance microwave fields.   
The electronic spin can be optically polarized and read out and has good coherence properties at ambient condition.  Phase acquired during the coupled evolution with the external nuclear spins is mapped into a population difference and  read out via spin-dependent fluorescent intensity. NV centers can be engineered in isotopically purified~\cite{Balasubramanian09} synthetic diamond by low-energy Nitrogen implantation~\cite{Romach14} or delta doping~\cite{Ohashi13,Ohno12}. These techniques provide good coherence properties and strong coupling to external spins. The ancillary nuclear spin is the intrinsic nuclear spin associated with the NV center, e.g., $^{15}$N spin-1/2 with hyperfine coupling of 3MHz.

\subsection{Hamiltonian Engineering by Dynamic Filter Generation}
The creation of a dynamic Bragg grating (spectral filter), which yields high spatial resolution, is obtained from the concatenation of the NV spin-lock Hamiltonian with a gradient field. The resulting evolution can be understood from a first order approximation~\cite{Suzuki76}, similar to average Hamiltonian theory~\cite{Haeberlen68}. Expanding the product in Eq.(~\ref{eq:evolution}) of the main text, we obtain 
{\small$
\left[e^{-i H_{G}t_g}e^{-iH_{\text{SL}}t/F}\right]^F \!\!=\! e^{-iH_Gt_gF}\prod_{f=0}^{F-1} e^{-i fH_{G}t_g}e^{iH_{\text{SL}}t/F}e^{if H_{G}t_g}.
$}\\
The gradient Hamiltonian only acts on the nuclear spins, rotating the spin-lock Hamiltonian to
{\small$
\widetilde H^{(f)}_{\text{SL}}=e^{-i fH_{G}t_g}H_{\text{SL}}e^{if H_Gt_g}$
$=\Omega S_z+(\omega_L+A^j)I^j_z +B^j_\perp (e^{ift_g(\omega_L+A^j)}e^{i\phi_j}  S_+I_-^j+\text{h.c.}).
$}\\
The product {\small$\prod_{f=0}^{F-1}e^{-i\widetilde H^{(f)}_{\text{SL}}t/F}$} can be approximated to first order by the exponential of the sum {\small$\approx\! e^{-it/F\sum \widetilde H^{(f)}_{\text{SL}}}$} and we thus obtained the effective Hamiltonian of Eq.~(\ref{eq:avHam}). The error in the approximation is given by the commutators {\small$[\widetilde H^{(f)}_{\text{SL}},\widetilde H^{(f')}_{\text{SL}}]t^2/F^2$ $\sim\omega_LB_\perp^jt^2/F^2$}. For typical values of the nuclear spin Larmor frequency and the transverse dipolar coupling, {\small$\omega_L\sim 100-1000 B_\perp^j$}, and setting {\small$t\sim1/B_\perp^j$}, we need {\small$F=10-30$} for the approximation to hold. This can be easily reached and it is usually required anyway to achieve the desired frequency selectivity.
Symmetrization of the sequence~\cite{Haeberlen68} achieves cancellation of all odd order terms making the filter insensitive to slow fluctuations in the microwave driving power at the cost of longer sequences (SI Appendix).

\subsection{Decoupling}
As the frequency selectivity of the NV sensing protocols increases, the spectral resolution is no longer limited by the filter itself but by the finite linewidth of the nuclear signal.
A major source of line broadening are the couplings among nuclear spins, which can be as large as tens of kHz between neighboring spins in a protein. 
Simple steps to mitigate this problems include reducing the spin density by  isotopic labeling~\cite{Kainosho06} and decoupling the spins of interest from different spin species (heteronuclear decoupling), using well-established methods in the NMR community~\cite{Haeberlen68}.
Contrary to DD-based sensing techniques, homonuclear decoupling sequences can be embedded in the cross-polarization scheme. Consider, e.g.,  one of the simplest decoupling sequences, WAHUHA~\cite{Waugh68}, which works by averaging to zero the secular homonuclear dipolar coupling over the cycle time. Assuming the sequence is applied during the NV driving, the effective Hamiltonian averaged over one cycle becomes\\
{\small$
H_{\text{SL}}\!=\!\Omega S_z+\textstyle\sum_j \left[(\omega_L+\frac{A^j}{2\sqrt3})I^j_{z_m}\!-\! B^j_m S_xI_{x_m}^j\!+\!iC^j S_xI_{y_m}^j\!+\!\text{h.c.})\right],$}
where  {\small$B^j_m=B^j_\perp\frac{\left(2+\sqrt{3}\right) \sin(\phi )+ \cos(\phi) }{3 \sqrt{2} \left(\sqrt{3}-3\right)}$}, 
{\small$C^j_m=B^j_\perp\frac{\left(2+\sqrt{3}\right) ( \cos(\phi )- \sin(\phi) }{3 \sqrt{2} \left(\sqrt{3}-3\right)}$} and the operators {\small$\vec I_m$} are defined in a frame where the nuclear spin z axis is rotated to the [111] direction. In this frame, the dynamics is equivalent to what observed in the absence of homonuclear decoupling; thus polarization is transferred from the NV electronic spin to the nuclear spin under the resonant condition, {\small$\Omega=\omega_L+A^j_m/(2\sqrt 3)$}, at a rate set by the transverse coupling, {\small$B^j_\perp\frac{\sqrt{7+4\sqrt3}}{3\sqrt{3}}$}. 
More generally, other decoupling sequences can be applied during the spin-locking periods (as well as  during evolution under the gradients in the filtered scheme). The sequences  narrow the nuclear spin linewidth, at the cost of a reduction in the dipolar coupling strength on the order of the homonuclear sequence scaling factor~\cite{Haeberlen68} (e.g. {\small$1/\sqrt3$} for WAHUHA).

\subsection{Volume Uncertainty Estimation}
The explicit functional dependence of the spins' reconstructed positions is intractable for any, but the fewest number of spins. Furthermore, if more coupling constants (both spin-NV and intra-spin) than spatial degrees of freedom are accessible, the system is overspecified and a unique solution compatible with all measured coupling will generally not exist, if the latter contain an error. To treat this problem in the simplest way, we perform a linear regression analysis around the true spatial positions and subsequently determine the error propagation within this framework, as described in detail the SI Appendix. The obtained volume error bound of the individual spins sets the figure of merit for our proposed sensing scheme. 

\end{materials}
 \begin{acknowledgments}
We thank F. Jelezko, E. Boyden, C. Belthangady, S. DeVience, I. Lovchinsky, L. Pham and A. Sushkov for discussion. This work was supported in part by the U.S. Army Research Office through a MURI grant No. W911NF-11-1-0400 and by DARPA (QuASAR program). 
\end{acknowledgments}

\bibliographystyle{pnas}

\small
\begin{thebibliography}{57}%
\makeatletter
\providecommand \@ifxundefined [1]{%
 \@ifx{#1\undefined}
}%
\providecommand \@ifnum [1]{%
 \ifnum #1\expandafter \@firstoftwo
 \else \expandafter \@secondoftwo
 \fi
}%
\providecommand \@ifx [1]{%
 \ifx #1\expandafter \@firstoftwo
 \else \expandafter \@secondoftwo
 \fi
}%
\providecommand \natexlab [1]{#1}%
\providecommand \enquote  [1]{``#1''}%
\providecommand \href@noop [0]{\@secondoftwo}%
\providecommand \href [0]{\begingroup \@sanitize@url \@href}%
\providecommand \@href[1]{\@@startlink{#1}\@@href}%
\providecommand \@@href[1]{\endgroup#1\@@endlink}%
\providecommand \@sanitize@url [0]{\catcode `\\12\catcode `\$12\catcode
  `\&12\catcode `\#12\catcode `\^12\catcode `\_12\catcode `\%12\relax}%
\providecommand \@@startlink[1]{}%
\providecommand \@@endlink[0]{}%
\providecommand \url  [0]{\begingroup\@sanitize@url \@url }%
\providecommand \@url [1]{\endgroup\@href {#1}{\urlprefix }}%
\providecommand \urlprefix  [0]{URL }%
\providecommand \Eprint [0]{\href }%
\providecommand \doibase [0]{http://dx.doi.org/}%
\providecommand \selectlanguage [0]{\@gobble}%
\providecommand \bibinfo  [0]{\@secondoftwo}%
\providecommand \bibfield  [0]{\@secondoftwo}%
\providecommand \translation [1]{[#1]}%
\providecommand \BibitemOpen [0]{}%
\providecommand \bibitemStop [0]{}%
\providecommand \bibitemNoStop [0]{.\EOS\space}%
\providecommand \EOS [0]{\spacefactor3000\relax}%
\providecommand \BibitemShut  [1]{\csname bibitem#1\endcsname}%
\let\auto@bib@innerbib\@empty
\bibitem [{ {Wimberly}\ \emph {et~al.}(2000)
  {Wimberly},  {Brodersen},  {William M.~Clemons},
   {Morgan-Warren},  {Carter}, 
  {Vonrhein},  {Hartsch},\ and\ 
  {Ramakrishnan}}]{Ramakrishnan00}%
  \BibitemOpen
  \bibfield  {author} {\bibinfo {author} { {B.~T.}\ 
  {Wimberly}}, \bibinfo {author} { {D.~E.}\ 
  {Brodersen}}, \bibinfo {author} { {J.}~ {William
  M.~Clemons}}, \bibinfo {author} { {R.~J.}\ 
  {Morgan-Warren}}, \bibinfo {author} { {A.~P.}\ 
  {Carter}}, \bibinfo {author} { {C.}~ {Vonrhein}},
  \bibinfo {author} { {T.}~ {Hartsch}}, \ and\
  \bibinfo {author} { {V.}~ {Ramakrishnan}},\
  }\href {\doibase 10.1038/35030006} {\bibfield  {journal} {\bibinfo  {journal} {Nature}\ }\textbf
  {\bibinfo {volume} {407}},\ \bibinfo {pages} {327} (\bibinfo {year}
  {2000})}\BibitemShut {NoStop}%
\bibitem [{ {Auer}(2000)}]{Auer00}%
  \BibitemOpen
  \bibfield  {author} {\bibinfo {author} { {M.~J.}\ 
  {Auer}},\ }\href@noop {} {\bibfield  {journal} {\bibinfo  {journal} {Mol.
  Med.}\ }\textbf {\bibinfo {volume} {78}},\ \bibinfo {pages} {191} (\bibinfo
  {year} {2000})}\BibitemShut {NoStop}%
\bibitem [{ {Wuthrich}(1986)}]{Wuthrich86}%
  \BibitemOpen
  \bibfield  {author} {\bibinfo {author} { {K.}~
  {Wuthrich}},\ }\href@noop {} {\emph {\bibinfo {title} {NMR of proteins and
  nucleic acids}}}\ (\bibinfo  {publisher} {Wiley},\ \bibinfo {year}
  {1986})\BibitemShut {NoStop}%
\bibitem [{ {Taylor}\ \emph {et~al.}(2008) {Taylor},
   {Cappellaro},  {Childress},  {Jiang},
   {Budker},  {Hemmer},  {Yacoby},
   {Walsworth},\ and\  {Lukin}}]{Taylor08}%
  \BibitemOpen
  \bibfield  {author} {\bibinfo {author} { {J.~M.}\ 
  {Taylor}}, \bibinfo {author} { {P.}~ {Cappellaro}},
  \bibinfo {author} { {L.}~ {Childress}}, \bibinfo
  {author} { {L.}~ {Jiang}}, \bibinfo {author}
  { {D.}~ {Budker}}, \bibinfo {author} {
  {P.~R.}\  {Hemmer}}, \bibinfo {author} {
  {A.}~ {Yacoby}}, \bibinfo {author} {
  {R.}~ {Walsworth}}, \ and\ \bibinfo {author} {
  {M.~D.}\  {Lukin}},\ }\href {\doibase 10.1038/nphys1075}
  {\bibfield  {journal} {\bibinfo  {journal} {Nat. Phys.}\ }\textbf {\bibinfo
  {volume} {4}},\ \bibinfo {pages} {810} (\bibinfo {year} {2008})}\BibitemShut
  {NoStop}%
\bibitem [{ {Maze}\ \emph {et~al.}(2008) {Maze},
   {Stanwix},  {Hodges},  {Hong},
   {Taylor},  {Cappellaro},  {Jiang},
   {Zibrov},  {Yacoby},  {Walsworth},\
  and\  {Lukin}}]{Maze08}%
  \BibitemOpen
  \bibfield  {author} {\bibinfo {author} { {J.~R.}\ 
  {Maze}}, \bibinfo {author} { {P.~L.}\  {Stanwix}},
  \bibinfo {author} { {J.~S.}\  {Hodges}}, \bibinfo
  {author} { {S.}~ {Hong}}, \bibinfo {author}
  { {J.~M.}\  {Taylor}}, \bibinfo {author}
  { {P.}~ {Cappellaro}}, \bibinfo {author}
  { {L.}~ {Jiang}}, \bibinfo {author} {
  {A.}~ {Zibrov}}, \bibinfo {author} {
  {A.}~ {Yacoby}}, \bibinfo {author} {
  {R.}~ {Walsworth}}, \ and\ \bibinfo {author} {
  {M.~D.}\  {Lukin}},\ }\href {\doibase 10.1038/nature07279}
  {\bibfield  {journal} {\bibinfo  {journal} {Nature}\ }\textbf {\bibinfo
  {volume} {455}},\ \bibinfo {pages} {644} (\bibinfo {year}
  {2008})}\BibitemShut {NoStop}%
\bibitem [{ {Balasubramanian}\ \emph {et~al.}(2008)
  {Balasubramanian},  {Chan},  {Kolesov},
   {Al-Hmoud},  {Tisler},  {Shin},
   {Kim},  {Wojcik},  {Hemmer},
   {Krueger},  {Hanke}, 
  {Leitenstorfer},  {Bratschitsch},  {Jelezko},\ and\
   {Wrachtrup}}]{Balasubramanian08}%
  \BibitemOpen
  \bibfield  {author} {\bibinfo {author} { {G.}~
  {Balasubramanian}}, \bibinfo {author} { {I.~Y.}\ 
  {Chan}}, \bibinfo {author} { {R.}~ {Kolesov}},
  \bibinfo {author} { {M.}~ {Al-Hmoud}}, \bibinfo
  {author} { {J.}~ {Tisler}}, \bibinfo {author}
  { {C.}~ {Shin}}, \bibinfo {author} {
  {C.}~ {Kim}}, \bibinfo {author} { {A.}~
  {Wojcik}}, \bibinfo {author} { {P.~R.}\  {Hemmer}},
  \bibinfo {author} { {A.}~ {Krueger}}, \bibinfo
  {author} { {T.}~ {Hanke}}, \bibinfo {author}
  { {A.}~ {Leitenstorfer}}, \bibinfo {author}
  { {R.}~ {Bratschitsch}}, \bibinfo {author}
  { {F.}~ {Jelezko}}, \ and\ \bibinfo {author}
  { {J.}~ {Wrachtrup}},\ }\href {\doibase
  10.1038/nature07278} {\bibfield  {journal} {\bibinfo  {journal} {Nature}\
  }\textbf {\bibinfo {volume} {455}},\ \bibinfo {pages} {648} (\bibinfo {year}
  {2008})}\BibitemShut {NoStop}%
\bibitem [{ {Cai}\ \emph {et~al.}(2013) {Cai},
   {Jelezko},  {Plenio},\ and\ 
  {Retzker}}]{Cai13b}%
  \BibitemOpen
  \bibfield  {author} {\bibinfo {author} { {J.}~
  {Cai}}, \bibinfo {author} { {F.}~ {Jelezko}},
  \bibinfo {author} { {M.~B.}\  {Plenio}}, \ and\
  \bibinfo {author} { {A.}~ {Retzker}},\ }\href
  {http://stacks.iop.org/1367-2630/15/i=1/a=013020} {\bibfield  {journal}
  {\bibinfo  {journal} {New J. Phys.}\ }\textbf {\bibinfo {volume} {15}},\
  \bibinfo {pages} {013020} (\bibinfo {year} {2013})}\BibitemShut {NoStop}%
\bibitem [{ {Hall}\ \emph {et~al.}(2012) {Hall},
   {Beart},  {Thomas},  {Simpson},
   {McGuinness},  {Cole},  {Manton},
   {Scholten},  {Jelezko},  {Wrachtrup},
   {Petrou},\ and\  {Hollenberg}}]{Hall12}%
  \BibitemOpen
  \bibfield  {author} {\bibinfo {author} { {L.~T.}\ 
  {Hall}}, \bibinfo {author} { {G.~C.~G.}\  {Beart}},
  \bibinfo {author} { {E.~A.}\  {Thomas}}, \bibinfo
  {author} { {D.~A.}\  {Simpson}}, \bibinfo {author}
  { {L.~P.}\  {McGuinness}}, \bibinfo {author}
  { {J.~H.}\  {Cole}}, \bibinfo {author}
  { {J.~H.}\  {Manton}}, \bibinfo {author}
  { {R.~E.}\  {Scholten}}, \bibinfo {author}
  { {F.}~ {Jelezko}}, \bibinfo {author} {
  {J.}~ {Wrachtrup}}, \bibinfo {author} {
  {S.}~ {Petrou}}, \ and\ \bibinfo {author} {
  {L.~C.~L.}\  {Hollenberg}},\ }\href {\doibase 10.1038/srep00401}
  {\bibfield  {journal} {\bibinfo  {journal} {Sci. Rep.}\ }\textbf {\bibinfo
  {volume} {2}},\ \bibinfo {pages} {401} (\bibinfo {year} {2012})}\BibitemShut
  {NoStop}%
\bibitem [{ {Cooper}\ \emph {et~al.}(2014) {Cooper},
   {Magesan},  {Yum},\ and\ 
  {Cappellaro}}]{Cooper14}%
  \BibitemOpen
  \bibfield  {author} {\bibinfo {author} { {A.}~
  {Cooper}}, \bibinfo {author} { {E.}~ {Magesan}},
  \bibinfo {author} { {H.}~ {Yum}}, \ and\ \bibinfo
  {author} { {P.}~ {Cappellaro}},\ }\href {\doibase
  10.1038/ncomms4141} {\bibfield  {journal} {\bibinfo  {journal} {Nat.
  Commun.}\ }\textbf {\bibinfo {volume} {5}},\ \bibinfo {pages} {3141}
  (\bibinfo {year} {2014})}\BibitemShut {NoStop}%
\bibitem [{ {{Le Sage}}\ \emph {et~al.}(2013) {{Le
  Sage}},  {Arai},  {Glenn}, 
  {DeVience},  {Pham},  {Rahn-Lee}, 
  {Lukin},  {Yacoby},  {Komeili},\ and\ 
  {Walsworth}}]{Lesage13}%
  \BibitemOpen
  \bibfield  {author} {\bibinfo {author} { {D.}~ {{Le
  Sage}}}, \bibinfo {author} { {K.}~ {Arai}}, \bibinfo
  {author} { {D.~R.}\  {Glenn}}, \bibinfo {author}
  { {S.~J.}\  {DeVience}}, \bibinfo {author}
  { {L.~M.}\  {Pham}}, \bibinfo {author}
  { {L.}~ {Rahn-Lee}}, \bibinfo {author}
  { {M.~D.}\  {Lukin}}, \bibinfo {author}
  { {A.}~ {Yacoby}}, \bibinfo {author} {
  {A.}~ {Komeili}}, \ and\ \bibinfo {author} {
  {R.~L.}\  {Walsworth}},\ }\href {\doibase 10.1038/nature12072}
  {\bibfield  {journal} {\bibinfo  {journal} {Nature}\ }\textbf {\bibinfo
  {volume} {496}},\ \bibinfo {pages} {486} (\bibinfo {year}
  {2013})}\BibitemShut {NoStop}%
\bibitem [{ {Taminiau}\ \emph {et~al.}(2012)
  {Taminiau},  {Wagenaar},  {van~der Sar},
   {Jelezko},  {Dobrovitski},\ and\ 
  {Hanson}}]{Taminiau12}%
  \BibitemOpen
  \bibfield  {author} {\bibinfo {author} { {T.~H.}\ 
  {Taminiau}}, \bibinfo {author} { {J.~J.~T.}\ 
  {Wagenaar}}, \bibinfo {author} { {T.}~ {van~der
  Sar}}, \bibinfo {author} { {F.}~ {Jelezko}},
  \bibinfo {author} { {V.~V.}\  {Dobrovitski}}, \ and\
  \bibinfo {author} { {R.}~ {Hanson}},\ }\href
  {\doibase 10.1103/PhysRevLett.109.137602} {\bibfield  {journal} {\bibinfo
  {journal} {Phys. Rev. Lett.}\ }\textbf {\bibinfo {volume} {109}},\ \bibinfo
  {pages} {137602} (\bibinfo {year} {2012})}\BibitemShut {NoStop}%
\bibitem [{ {Kolkowitz}\ \emph {et~al.}(2012)
  {Kolkowitz},  {Unterreithmeier},  {Bennett},\ and\
   {Lukin}}]{Kolkowitz12a}%
  \BibitemOpen
  \bibfield  {author} {\bibinfo {author} { {S.}~
  {Kolkowitz}}, \bibinfo {author} { {Q.~P.}\ 
  {Unterreithmeier}}, \bibinfo {author} { {S.~D.}\ 
  {Bennett}}, \ and\ \bibinfo {author} { {M.~D.}\ 
  {Lukin}},\ }\href {\doibase 10.1103/PhysRevLett.109.137601} {\bibfield
  {journal} {\bibinfo  {journal} {Phys. Rev. Lett.}\ }\textbf {\bibinfo
  {volume} {109}},\ \bibinfo {pages} {137601} (\bibinfo {year}
  {2012})}\BibitemShut {NoStop}%
\bibitem [{ {Zhao}\ \emph {et~al.}(2012) {Zhao},
   {Honert},  {Schmid},  {Klas},
   {Isoya},  {Markham},  {Twitchen},
   {Jelezko},  {Liu},  {Fedder},\ and\
   {Wrachtrup}}]{Zhao12}%
  \BibitemOpen
  \bibfield  {author} {\bibinfo {author} { {N.}~
  {Zhao}}, \bibinfo {author} { {J.}~ {Honert}},
  \bibinfo {author} { {B.}~ {Schmid}}, \bibinfo
  {author} { {M.}~ {Klas}}, \bibinfo {author}
  { {J.}~ {Isoya}}, \bibinfo {author} {
  {M.}~ {Markham}}, \bibinfo {author} {
  {D.}~ {Twitchen}}, \bibinfo {author} {
  {F.}~ {Jelezko}}, \bibinfo {author} { {R.-B.}\
   {Liu}}, \bibinfo {author} { {H.}~
  {Fedder}}, \ and\ \bibinfo {author} { {J.}~
  {Wrachtrup}},\ }\href {\doibase 10.1038/nnano.2012.152} {\bibfield  {journal}
  {\bibinfo  {journal} {Nat. Nanotech.}\ }\textbf {\bibinfo {volume} {7}},\
  \bibinfo {pages} {657} (\bibinfo {year} {2012})}\BibitemShut {NoStop}%
\bibitem [{ {Mamin}\ \emph {et~al.}(2013) {Mamin},
   {Kim},  {Sherwood},  {Rettner},
   {Ohno},  {Awschalom},\ and\ 
  {Rugar}}]{Mamin13}%
  \BibitemOpen
  \bibfield  {author} {\bibinfo {author} { {H.~J.}\ 
  {Mamin}}, \bibinfo {author} { {M.}~ {Kim}}, \bibinfo
  {author} { {M.~H.}\  {Sherwood}}, \bibinfo {author}
  { {C.~T.}\  {Rettner}}, \bibinfo {author}
  { {K.}~ {Ohno}}, \bibinfo {author} {
  {D.~D.}\  {Awschalom}}, \ and\ \bibinfo {author} {
  {D.}~ {Rugar}},\ }\href {\doibase 10.1126/science.1231540}
  {\bibfield  {journal} {\bibinfo  {journal} {Science}\ }\textbf {\bibinfo
  {volume} {339}},\ \bibinfo {pages} {557} (\bibinfo {year}
  {2013})}\BibitemShut {NoStop}%
\bibitem [{ {Staudacher}\ \emph {et~al.}(2013)
  {Staudacher},  {Shi},  {Pezzagna}, 
  {Meijer},  {Du},  {Meriles}, 
  {Reinhard},\ and\  {Wrachtrup}}]{Staudacher13}%
  \BibitemOpen
  \bibfield  {author} {\bibinfo {author} { {T.}~
  {Staudacher}}, \bibinfo {author} { {F.}~ {Shi}},
  \bibinfo {author} { {S.}~ {Pezzagna}}, \bibinfo
  {author} { {J.}~ {Meijer}}, \bibinfo {author}
  { {J.}~ {Du}}, \bibinfo {author} {
  {C.~A.}\  {Meriles}}, \bibinfo {author} {
  {F.}~ {Reinhard}}, \ and\ \bibinfo {author} {
  {J.}~ {Wrachtrup}},\ }\href {\doibase 10.1126/science.1231675}
  {\bibfield  {journal} {\bibinfo  {journal} {Science}\ }\textbf {\bibinfo
  {volume} {339}},\ \bibinfo {pages} {561} (\bibinfo {year}
  {2013})}\BibitemShut {NoStop}%
\bibitem [{ {Ohashi}\ \emph {et~al.}(2013) {Ohashi},
   {Rosskopf},  {Watanabe},  {Loretz},
   {Tao},  {Hauert},  {Tomizawa},
   {Ishikawa},  {Ishi-Hayase}, 
  {Shikata},  {Degen},\ and\  {Itoh}}]{Ohashi13}%
  \BibitemOpen
  \bibfield  {author} {\bibinfo {author} { {K.}~
  {Ohashi}}, \bibinfo {author} { {T.}~ {Rosskopf}},
  \bibinfo {author} { {H.}~ {Watanabe}}, \bibinfo
  {author} { {M.}~ {Loretz}}, \bibinfo {author}
  { {Y.}~ {Tao}}, \bibinfo {author} {
  {R.}~ {Hauert}}, \bibinfo {author} {
  {S.}~ {Tomizawa}}, \bibinfo {author} {
  {T.}~ {Ishikawa}}, \bibinfo {author} {
  {J.}~ {Ishi-Hayase}}, \bibinfo {author} {
  {S.}~ {Shikata}}, \bibinfo {author} { {C.~L.}\
   {Degen}}, \ and\ \bibinfo {author} { {K.~M.}\
   {Itoh}},\ }\href {\doibase 10.1021/nl402286v} {\bibfield
  {journal} {\bibinfo  {journal} {Nano Letters}\ }\textbf {\bibinfo {volume}
  {13}},\ \bibinfo {pages} {4733} (\bibinfo {year} {2013})}\BibitemShut
  {NoStop}%
\bibitem [{ {{Mamin}}\ \emph {et~al.}(2014)
  {{Mamin}},  {{Sherwood}},  {{Kim}}, 
  {{Rettner}},  {{Ohno}},  {{Awschalom}},\ and\
   {{Rugar}}}]{Mamin14}%
  \BibitemOpen
  \bibfield  {author} {\bibinfo {author} { {H.~J.}\ 
  {{Mamin}}}, \bibinfo {author} { {M.~H.}\ 
  {{Sherwood}}}, \bibinfo {author} { {M.}~ {{Kim}}},
  \bibinfo {author} { {C.~T.}\  {{Rettner}}}, \bibinfo
  {author} { {K.}~ {{Ohno}}}, \bibinfo {author}
  { {D.~D.}\  {{Awschalom}}}, \ and\ \bibinfo {author}
  { {D.}~ {{Rugar}}},\ }\href@noop {} {\bibfield
  {journal} {\bibinfo  {journal} {ArXiv e-prints}\ } (\bibinfo {year}
  {2014})},\ \Eprint {http://arxiv.org/abs/1404.7480} {arXiv:1404.7480} \BibitemShut {NoStop}%
\bibitem [{ {Grinolds}\ \emph {et~al.}(2014)
  {Grinolds},  {Warner},  {De~Greve}, 
  {Dovzhenko},  {Thiel},  {Walsworth}, 
  {Hong},  {Maletinsky},\ and\ 
  {Yacoby}}]{Grinolds14}%
  \BibitemOpen
  \bibfield  {author} {\bibinfo {author} { {M.}~
  {Grinolds}}, \bibinfo {author} { {M.}~ {Warner}},
  \bibinfo {author} { {K.}~ {De~Greve}}, \bibinfo
  {author} { {Y.}~ {Dovzhenko}}, \bibinfo {author}
  { {L.}~ {Thiel}}, \bibinfo {author} {
  {R.}~ {Walsworth}}, \bibinfo {author} {
  {S.}~ {Hong}}, \bibinfo {author} { {P.}~
  {Maletinsky}}, \ and\ \bibinfo {author} { {A.}~
  {Yacoby}},\ }\href {http://dx.doi.org/10.1038/nnano.2014.30} {\bibfield
  {journal} {\bibinfo  {journal} {Nat. Nanotech.}\ }\textbf {\bibinfo {volume}
  {9}},\ \bibinfo {pages} {279} (\bibinfo {year} {2014})}\BibitemShut {NoStop}%
\bibitem [{ {Ermakova}\ \emph {et~al.}(2013)
  {Ermakova},  {Pramanik},  {Cai}, 
  {Algara-Siller},  {Kaiser},  {Weil}, 
  {Tzeng},  {Chang},  {McGuinness}, 
  {Plenio},  {Naydenov},\ and\ 
  {Jelezko}}]{Ermakova13}%
  \BibitemOpen
  \bibfield  {author} {\bibinfo {author} { {A.}~
  {Ermakova}}, \bibinfo {author} { {G.}~ {Pramanik}},
  \bibinfo {author} { {J.-M.}\  {Cai}}, \bibinfo
  {author} { {G.}~ {Algara-Siller}}, \bibinfo {author}
  { {U.}~ {Kaiser}}, \bibinfo {author} {
  {T.}~ {Weil}}, \bibinfo {author} { {Y.-K.}\
   {Tzeng}}, \bibinfo {author} { {H.~C.}\ 
  {Chang}}, \bibinfo {author} { {L.~P.}\ 
  {McGuinness}}, \bibinfo {author} { {M.~B.}\ 
  {Plenio}}, \bibinfo {author} { {B.}~ {Naydenov}}, \
  and\ \bibinfo {author} { {F.}~ {Jelezko}},\ }\href
  {\doibase 10.1021/nl4015233} {\bibfield  {journal} {\bibinfo  {journal} {Nano
  Letters}\ }\textbf {\bibinfo {volume} {13}},\ \bibinfo {pages} {3305}
  (\bibinfo {year} {2013})}\BibitemShut {NoStop}%
\bibitem [{ {Kaufmann}\ \emph {et~al.}(2013)
  {Kaufmann},  {Simpson},  {Hall}, 
  {Perunicic},  {Senn},  {Steinert}, 
  {McGuinness},  {Johnson},  {Ohshima}, 
  {Caruso},  {Wrachtrup},  {Scholten}, 
  {Mulvaney},\ and\  {Hollenberg}}]{Kaufmann13}%
  \BibitemOpen
  \bibfield  {author} {\bibinfo {author} { {S.}~
  {Kaufmann}}, \bibinfo {author} { {D.~A.}\ 
  {Simpson}}, \bibinfo {author} { {L.~T.}\  {Hall}},
  \bibinfo {author} { {V.}~ {Perunicic}}, \bibinfo
  {author} { {P.}~ {Senn}}, \bibinfo {author}
  { {S.}~ {Steinert}}, \bibinfo {author}
  { {L.~P.}\  {McGuinness}}, \bibinfo {author}
  { {B.~C.}\  {Johnson}}, \bibinfo {author}
  { {T.}~ {Ohshima}}, \bibinfo {author} {
  {F.}~ {Caruso}}, \bibinfo {author} {
  {J.}~ {Wrachtrup}}, \bibinfo {author} { {R.~E.}\
   {Scholten}}, \bibinfo {author} { {P.}~
  {Mulvaney}}, \ and\ \bibinfo {author} { {L.}~
  {Hollenberg}},\ }\href {\doibase 10.1073/pnas.1300640110} {\bibfield
  {journal} {\bibinfo  {journal} {Proc. Nat Acad. Sc.}\ }\textbf {\bibinfo
  {volume} {110}},\ \bibinfo {pages} {10894} (\bibinfo {year}
  {2013})}\BibitemShut {NoStop}%
\bibitem [{ {{Sushkov}}\ \emph {et~al.}(2013)
  {{Sushkov}},  {{Chisholm}},  {{Lovchinsky}},
   {{Kubo}},  {{Lo}},  {{Bennett}},
   {{Hunger}},  {{Akimov}}, 
  {{Walsworth}},  {{Park}},\ and\ 
  {{Lukin}}}]{Sushkov13u}%
  \BibitemOpen
  \bibfield  {author} {\bibinfo {author} { {A.~O.}\ 
  {{Sushkov}}}, \bibinfo {author} { {N.}~
  {{Chisholm}}}, \bibinfo {author} { {I.}~
  {{Lovchinsky}}}, \bibinfo {author} { {M.}~
  {{Kubo}}}, \bibinfo {author} { {P.~K.}\  {{Lo}}},
  \bibinfo {author} { {S.~D.}\  {{Bennett}}}, \bibinfo
  {author} { {D.}~ {{Hunger}}}, \bibinfo {author}
  { {A.}~ {{Akimov}}}, \bibinfo {author}
  { {R.~L.}\  {{Walsworth}}}, \bibinfo {author}
  { {H.}~ {{Park}}}, \ and\ \bibinfo {author}
  { {M.~D.}\  {{Lukin}}},\ }\href@noop {} {\bibfield
  {journal} {\bibinfo  {journal} {ArXiv e-prints},\ \bibinfo {pages}
 \Eprint {http://arxiv.org/abs/1311.1801} {arXiv:1311.1801}  } (\bibinfo {year} {2013})}\BibitemShut {NoStop}%
\bibitem [{ {Balasubramanian}\ \emph {et~al.}(2009)
  {Balasubramanian},  {Neumann},  {Twitchen},
   {Markham},  {Kolesov},  {Mizuochi},
   {Isoya},  {Achard},  {Beck},
   {Tissler},  {Jacques},  {Hemmer},
   {Jelezko},\ and\ 
  {Wrachtrup}}]{Balasubramanian09}%
  \BibitemOpen
  \bibfield  {author} {\bibinfo {author} { {G.}~
  {Balasubramanian}}, \bibinfo {author} { {P.}~
  {Neumann}}, \bibinfo {author} { {D.}~ {Twitchen}},
  \bibinfo {author} { {M.}~ {Markham}}, \bibinfo
  {author} { {R.}~ {Kolesov}}, \bibinfo {author}
  { {N.}~ {Mizuochi}}, \bibinfo {author}
  { {J.}~ {Isoya}}, \bibinfo {author} {
  {J.}~ {Achard}}, \bibinfo {author} {
  {J.}~ {Beck}}, \bibinfo {author} { {J.}~
  {Tissler}}, \bibinfo {author} { {V.}~ {Jacques}},
  \bibinfo {author} { {P.~R.}\  {Hemmer}}, \bibinfo
  {author} { {F.}~ {Jelezko}}, \ and\ \bibinfo
  {author} { {J.}~ {Wrachtrup}},\ }\href {\doibase
  10.1038/nmat2420} {\bibfield  {journal} {\bibinfo  {journal} {Nat. Mater.}\
  }\textbf {\bibinfo {volume} {8}},\ \bibinfo {pages} {383} (\bibinfo {year}
  {2009})}\BibitemShut {NoStop}%
\bibitem [{ {Staudacher}\ \emph {et~al.}(2012)
  {Staudacher},  {Ziem},  {H\"{a}ussler},
   {St\"{o}hr},  {Steinert}, 
  {Reinhard},  {Scharpf},  {Denisenko},\ and\
   {Wrachtrup}}]{Staudacher12}%
  \BibitemOpen
  \bibfield  {author} {\bibinfo {author} { {T.}~
  {Staudacher}}, \bibinfo {author} { {F.}~ {Ziem}},
  \bibinfo {author} { {L.}~ {H\"{a}ussler}}, \bibinfo
  {author} { {R.}~ {St\"{o}hr}}, \bibinfo {author}
  { {S.}~ {Steinert}}, \bibinfo {author}
  { {F.}~ {Reinhard}}, \bibinfo {author}
  { {J.}~ {Scharpf}}, \bibinfo {author} {
  {A.}~ {Denisenko}}, \ and\ \bibinfo {author} {
  {J.}~ {Wrachtrup}},\ }\href {\doibase 10.1063/1.4767144}
  {\bibfield  {journal} {\bibinfo  {journal} {Appl. Phys. Lett.}\ }\textbf
  {\bibinfo {volume} {101}},\ \bibinfo {eid} {212401} (\bibinfo {year}
  {2012})}\BibitemShut {NoStop}%
\bibitem [{ {Hirose}\ \emph {et~al.}(2012) {Hirose},
   {Aiello},\ and\  {Cappellaro}}]{Hirose12}%
  \BibitemOpen
  \bibfield  {author} {\bibinfo {author} { {M.}~
  {Hirose}}, \bibinfo {author} { {C.~D.}\  {Aiello}},
  \ and\ \bibinfo {author} { {P.}~ {Cappellaro}},\
  }\href {\doibase 10.1103/PhysRevA.86.062320} {\bibfield  {journal} {\bibinfo
  {journal} {Phys. Rev. A}\ }\textbf {\bibinfo {volume} {86}},\ \bibinfo
  {pages} {062320} (\bibinfo {year} {2012})}\BibitemShut {NoStop}%
\bibitem [{ {Loretz}\ \emph {et~al.}(2013) {Loretz},
   {Rosskopf},\ and\  {Degen}}]{Loretz13}%
  \BibitemOpen
  \bibfield  {author} {\bibinfo {author} { {M.}~
  {Loretz}}, \bibinfo {author} { {T.}~ {Rosskopf}}, \
  and\ \bibinfo {author} { {C.~L.}\  {Degen}},\ }\href
  {\doibase 10.1103/PhysRevLett.110.017602} {\bibfield  {journal} {\bibinfo
  {journal} {Phys. Rev. Lett.}\ }\textbf {\bibinfo {volume} {110}},\ \bibinfo
  {pages} {017602} (\bibinfo {year} {2013})}\BibitemShut {NoStop}%
\bibitem [{ {Wu}\ \emph {et~al.}(2010) {Wu},
   {Chien},  {Mol},  {Fenalti},
   {Liu},  {Katritch},  {Abagyan},
   {Brooun},  {Wells},  {Bi},
   {Hamel},  {Kuhn},  {Handel},
   {Cherezov},\ and\  {Stevens}}]{Wu10}%
  \BibitemOpen
  \bibfield  {author} {\bibinfo {author} { {B.}~
  {Wu}}, \bibinfo {author} { {E.~Y.~T.}\  {Chien}},
  \bibinfo {author} { {C.~D.}\  {Mol}}, \bibinfo
  {author} { {G.}~ {Fenalti}}, \bibinfo {author}
  { {W.}~ {Liu}}, \bibinfo {author} {
  {V.}~ {Katritch}}, \bibinfo {author} {
  {R.}~ {Abagyan}}, \bibinfo {author} {
  {A.}~ {Brooun}}, \bibinfo {author} {
  {P.}~ {Wells}}, \bibinfo {author} { {F.~C.}\
   {Bi}}, \bibinfo {author} { {D.~J.}\ 
  {Hamel}}, \bibinfo {author} { {P.}~ {Kuhn}},
  \bibinfo {author} { {T.~M.}\  {Handel}}, \bibinfo
  {author} { {V.}~ {Cherezov}}, \ and\ \bibinfo
  {author} { {R.~C.}\  {Stevens}},\ }\href {\doibase
  10.1126/science.1194396} {\bibfield  {journal} {\bibinfo  {journal}
  {Science}\ }\textbf {\bibinfo {volume} {330}},\ \bibinfo {pages} {1066}
  (\bibinfo {year} {2010})}\BibitemShut {NoStop}%
\bibitem [{ {Ajoy}\ and\ 
  {Cappellaro}(2013)}]{Ajoy13l}%
  \BibitemOpen
  \bibfield  {author} {\bibinfo {author} { {A.}~
  {Ajoy}}\ and\ \bibinfo {author} { {P.}~
  {Cappellaro}},\ }\href {\doibase 10.1103/PhysRevLett.110.220503} {\bibfield
  {journal} {\bibinfo  {journal} {Phys. Rev. Lett.}\ }\textbf {\bibinfo
  {volume} {110}},\ \bibinfo {pages} {220503} (\bibinfo {year}
  {2013})}\BibitemShut {NoStop}%
\bibitem [{ {Meiboom}\ and\ 
  {Gill}(1958)}]{Meiboom58}%
  \BibitemOpen
  \bibfield  {author} {\bibinfo {author} { {S.}~
  {Meiboom}}\ and\ \bibinfo {author} { {D.}~ {Gill}},\
  }\href {\doibase 10.1063/1.1716296} {\bibfield  {journal} {\bibinfo
  {journal} {Rev. Sc. Instr.}\ }\textbf {\bibinfo {volume} {29}},\ \bibinfo
  {pages} {688} (\bibinfo {year} {1958})}\BibitemShut {NoStop}%
\bibitem [{ {Biercuk}\ \emph {et~al.}(2011)
  {Biercuk},  {Doherty},\ and\  {Uys}}]{Biercuk11}%
  \BibitemOpen
  \bibfield  {author} {\bibinfo {author} { {M.~J.}\ 
  {Biercuk}}, \bibinfo {author} { {A.~C.}\ 
  {Doherty}}, \ and\ \bibinfo {author} { {H.}~
  {Uys}},\ }\href {\doibase 10.1088/0953-4075/44/15/154002} {\bibfield
  {journal} {\bibinfo  {journal} {J. of Phys. B}\ }\textbf {\bibinfo {volume}
  {44}},\ \bibinfo {pages} {154002} (\bibinfo {year} {2011})}\BibitemShut
  {NoStop}%
\bibitem [{ {Ajoy}\ \emph {et~al.}(2011) {Ajoy},
   {\'Alvarez},\ and\  {Suter}}]{Ajoy11}%
  \BibitemOpen
  \bibfield  {author} {\bibinfo {author} { {A.}~
  {Ajoy}}, \bibinfo {author} { {G.~A.}\  {\'Alvarez}},
  \ and\ \bibinfo {author} { {D.}~ {Suter}},\ }\href
  {\doibase 10.1103/PhysRevA.83.032303} {\bibfield  {journal} {\bibinfo
  {journal} {Phys. Rev. A}\ }\textbf {\bibinfo {volume} {83}},\ \bibinfo
  {pages} {032303} (\bibinfo {year} {2011})}\BibitemShut {NoStop}%
\bibitem [{ {London}\ \emph {et~al.}(2013) {London},
   {Scheuer},  {Cai},  {Schwarz},
   {Retzker},  {Plenio},  {Katagiri},
   {Teraji},  {Koizumi},  {Isoya},
   {Fischer},  {McGuinness}, 
  {Naydenov},\ and\  {Jelezko}}]{London13}%
  \BibitemOpen
  \bibfield  {author} {\bibinfo {author} { {P.}~
  {London}}, \bibinfo {author} { {J.}~ {Scheuer}},
  \bibinfo {author} { {J.-M.}\  {Cai}}, \bibinfo
  {author} { {I.}~ {Schwarz}}, \bibinfo {author}
  { {A.}~ {Retzker}}, \bibinfo {author} {
  {M.~B.}\  {Plenio}}, \bibinfo {author} {
  {M.}~ {Katagiri}}, \bibinfo {author} {
  {T.}~ {Teraji}}, \bibinfo {author} {
  {S.}~ {Koizumi}}, \bibinfo {author} {
  {J.}~ {Isoya}}, \bibinfo {author} {
  {R.}~ {Fischer}}, \bibinfo {author} { {L.~P.}\
   {McGuinness}}, \bibinfo {author} {
  {B.}~ {Naydenov}}, \ and\ \bibinfo {author} {
  {F.}~ {Jelezko}},\ }\href {\doibase
  10.1103/PhysRevLett.111.067601} {\bibfield  {journal} {\bibinfo  {journal}
  {Phys. Rev. Lett.}\ }\textbf {\bibinfo {volume} {111}},\ \bibinfo {pages}
  {067601} (\bibinfo {year} {2013})}\BibitemShut {NoStop}%
\bibitem [{ {Belthangady}\ \emph {et~al.}(2013)
  {Belthangady},  {Bar-Gill},  {Pham}, 
  {Arai},  {Le~Sage},  {Cappellaro},\ and\
   {Walsworth}}]{Belthangady13}%
  \BibitemOpen
  \bibfield  {author} {\bibinfo {author} { {C.}~
  {Belthangady}}, \bibinfo {author} { {N.}~
  {Bar-Gill}}, \bibinfo {author} { {L.~M.}\  {Pham}},
  \bibinfo {author} { {K.}~ {Arai}}, \bibinfo {author}
  { {D.}~ {Le~Sage}}, \bibinfo {author} {
  {P.}~ {Cappellaro}}, \ and\ \bibinfo {author} {
  {R.~L.}\  {Walsworth}},\ }\href {\doibase
  10.1103/PhysRevLett.110.157601} {\bibfield  {journal} {\bibinfo  {journal}
  {Phys. Rev. Lett.}\ }\textbf {\bibinfo {volume} {110}},\ \bibinfo {pages}
  {157601} (\bibinfo {year} {2013})}\BibitemShut {NoStop}%
\bibitem [{ {Hartmann}\ and\ 
  {Hahn}(1962)}]{Hartmann62}%
  \BibitemOpen
  \bibfield  {author} {\bibinfo {author} { {S.~R.}\ 
  {Hartmann}}\ and\ \bibinfo {author} { {E.~L.}\ 
  {Hahn}},\ }\href {\doibase 10.1103/PhysRev.128.2042} {\bibfield  {journal}
  {\bibinfo  {journal} {Phys. Rev.}\ }\textbf {\bibinfo {volume} {128}},\
  \bibinfo {pages} {2042} (\bibinfo {year} {1962})}\BibitemShut {NoStop}%
\bibitem [{ {Henstra}\ \emph {et~al.}(1988)
  {Henstra},  {Dirksen},  {Schmidt},\ and\
   {Wenckebach}}]{Henstra88}%
  \BibitemOpen
  \bibfield  {author} {\bibinfo {author} { {A.}~
  {Henstra}}, \bibinfo {author} { {P.}~ {Dirksen}},
  \bibinfo {author} { {J.}~ {Schmidt}}, \ and\
  \bibinfo {author} { {W.}~ {Wenckebach}},\ }\href
  {\doibase 10.1016/0022-2364(88)90190-4} {\bibfield  {journal} {\bibinfo
  {journal} {J. Mag. Res.}\ }\textbf {\bibinfo {volume} {77}},\ \bibinfo
  {pages} {389} (\bibinfo {year} {1988})}\BibitemShut {NoStop}%
\bibitem [{ {Degen}\ \emph {et~al.}(2009) {Degen},
   {Poggio},  {Mamin},  {Rettner},\ and\
   {Rugar}}]{Degen09}%
  \BibitemOpen
  \bibfield  {author} {\bibinfo {author} { {C.~L.}\ 
  {Degen}}, \bibinfo {author} { {M.}~ {Poggio}},
  \bibinfo {author} { {H.~J.}\  {Mamin}}, \bibinfo
  {author} { {C.~T.}\  {Rettner}}, \ and\ \bibinfo
  {author} { {D.}~ {Rugar}},\ }\href {\doibase
  10.1073/pnas.0812068106} {\bibfield  {journal} {\bibinfo  {journal} {Proc.
  Nat Acad. Sc.}\ }\textbf {\bibinfo {volume} {106}},\ \bibinfo {pages} {1313}
  (\bibinfo {year} {2009})}\BibitemShut {NoStop}%
\bibitem [{ {Arai}\ \emph {et~al.}(2013) {Arai},
   {Le~Sage},  {DeVience},  {Glenn},
   {Pham},  {Rahn-Lee},  {Lukin},
   {Yacoby},  {Komeili},\ and\ 
  {Walsworth}}]{Arai13}%
  \BibitemOpen
  \bibfield  {author} {\bibinfo {author} { {K.}~
  {Arai}}, \bibinfo {author} { {D.}~ {Le~Sage}},
  \bibinfo {author} { {S.~J.}\  {DeVience}}, \bibinfo
  {author} { {D.~R.}\  {Glenn}}, \bibinfo {author}
  { {L.~M.}\  {Pham}}, \bibinfo {author}
  { {L.}~ {Rahn-Lee}}, \bibinfo {author}
  { {M.~D.}\  {Lukin}}, \bibinfo {author}
  { {A.}~ {Yacoby}}, \bibinfo {author} {
  {A.}~ {Komeili}}, \ and\ \bibinfo {author} {
  {R.~L.}\  {Walsworth}},\ }\href
  {http://www.sciencedirect.com/science/article/pii/S000634951202334X}
  {\bibfield  {journal} {\bibinfo  {journal} {Biophysical Journal}\ }\textbf
  {\bibinfo {volume} {104}},\ \bibinfo {pages} {193a} (\bibinfo {year}
  {2013})}\BibitemShut {NoStop}%
\bibitem [{ {Schaffry}\ \emph {et~al.}(2011)
  {Schaffry},  {Gauger},  {Morton},\ and\
   {Benjamin}}]{Schaffry11}%
  \BibitemOpen
  \bibfield  {author} {\bibinfo {author} { {M.}~
  {Schaffry}}, \bibinfo {author} { {E.~M.}\ 
  {Gauger}}, \bibinfo {author} { {J.~J.~L.}\ 
  {Morton}}, \ and\ \bibinfo {author} { {S.~C.}\ 
  {Benjamin}},\ }\href {\doibase 10.1103/PhysRevLett.107.207210} {\bibfield
  {journal} {\bibinfo  {journal} {Phys. Rev. Lett.}\ }\textbf {\bibinfo
  {volume} {107}},\ \bibinfo {pages} {207210} (\bibinfo {year}
  {2011})}\BibitemShut {NoStop}%
\bibitem [{ {Nielsen}\ and\ 
  {Chuang}(2000)}]{Nielsen00b}%
  \BibitemOpen
  \bibfield  {author} {\bibinfo {author} { {M.~A.}\ 
  {Nielsen}}\ and\ \bibinfo {author} { {I.~L.}\ 
  {Chuang}},\ }\href@noop {} {\emph {\bibinfo {title} {Quantum computation and
  quantum information}}}\ (\bibinfo  {publisher} {Cambridge University Press},\
  \bibinfo {address} {Cambridge; New York},\ \bibinfo {year}
  {2000})\BibitemShut {NoStop}%
\bibitem [{ {Jiang}\ \emph {et~al.}(2009) {Jiang},
   {Hodges},  {Maze},  {Maurer},
   {Taylor},  {Cory},  {Hemmer},
   {Walsworth},  {Yacoby},  {Zibrov},\
  and\  {Lukin}}]{Jiang09}%
  \BibitemOpen
  \bibfield  {author} {\bibinfo {author} { {L.}~
  {Jiang}}, \bibinfo {author} { {J.~S.}\  {Hodges}},
  \bibinfo {author} { {J.~R.}\  {Maze}}, \bibinfo
  {author} { {P.}~ {Maurer}}, \bibinfo {author}
  { {J.~M.}\  {Taylor}}, \bibinfo {author}
  { {D.~G.}\  {Cory}}, \bibinfo {author}
  { {P.~R.}\  {Hemmer}}, \bibinfo {author}
  { {R.~L.}\  {Walsworth}}, \bibinfo {author}
  { {A.}~ {Yacoby}}, \bibinfo {author} {
  {A.~S.}\  {Zibrov}}, \ and\ \bibinfo {author} {
  {M.~D.}\  {Lukin}},\ }\href {\doibase 10.1126/science.1176496}
  {\bibfield  {journal} {\bibinfo  {journal} {Science}\ }\textbf {\bibinfo
  {volume} {326}},\ \bibinfo {pages} {267} (\bibinfo {year}
  {2009})}\BibitemShut {NoStop}%
\bibitem [{ {Cappellaro}\ \emph {et~al.}(2009)
  {Cappellaro},  {Jiang},  {Hodges},\ and\
   {Lukin}}]{Cappellaro09}%
  \BibitemOpen
  \bibfield  {author} {\bibinfo {author} { {P.}~
  {Cappellaro}}, \bibinfo {author} { {L.}~ {Jiang}},
  \bibinfo {author} { {J.~S.}\  {Hodges}}, \ and\
  \bibinfo {author} { {M.~D.}\  {Lukin}},\ }\href
  {\doibase 10.1103/PhysRevLett.102.210502} {\bibfield  {journal} {\bibinfo
  {journal} {Phys. Rev. Lett.}\ }\textbf {\bibinfo {volume} {102}},\ \bibinfo
  {eid} {210502} (\bibinfo {year} {2009})}\BibitemShut {NoStop}%
\bibitem [{ {van~der Sar}\ \emph {et~al.}(2012)
  {van~der Sar},  {Wang},  {Blok}, 
  {Bernien},  {Taminiau},  {Toyli}, 
  {Lidar},  {Awschalom},  {Hanson},\ and\
   {Dobrovitski}}]{van12}%
  \BibitemOpen
  \bibfield  {author} {\bibinfo {author} { {T.}~
  {van~der Sar}}, \bibinfo {author} { {Z.}~ {Wang}},
  \bibinfo {author} { {M.}~ {Blok}}, \bibinfo {author}
  { {H.}~ {Bernien}}, \bibinfo {author} {
  {T.}~ {Taminiau}}, \bibinfo {author} {
  {D.}~ {Toyli}}, \bibinfo {author} {
  {D.}~ {Lidar}}, \bibinfo {author} {
  {D.}~ {Awschalom}}, \bibinfo {author} {
  {R.}~ {Hanson}}, \ and\ \bibinfo {author} {
  {V.}~ {Dobrovitski}},\ }\href@noop {} {\bibfield  {journal}
  {\bibinfo  {journal} {Nature}\ }\textbf {\bibinfo {volume} {484}},\ \bibinfo
  {pages} {82} (\bibinfo {year} {2012})}\BibitemShut {NoStop}%
\bibitem [{ {Waugh}\ \emph {et~al.}(1968) {Waugh},
   {Huber},\ and\  {Haeberlen}}]{Waugh68}%
  \BibitemOpen
  \bibfield  {author} {\bibinfo {author} { {J.}~
  {Waugh}}, \bibinfo {author} { {L.}~ {Huber}}, \ and\
  \bibinfo {author} { {U.}~ {Haeberlen}},\ }\href
  {\doibase 10.1103/PhysRevLett.20.180} {\bibfield  {journal} {\bibinfo
  {journal} {Phys. Rev. Lett.}\ }\textbf {\bibinfo {volume} {20}},\ \bibinfo
  {pages} {180} (\bibinfo {year} {1968})}\BibitemShut {NoStop}%
\bibitem [{ {Khutsishvili}(1966)}]{Khutsishvili66}%
  \BibitemOpen
  \bibfield  {author} {\bibinfo {author} { {G.~R.}\ 
  {Khutsishvili}},\ }\href@noop {} {\bibfield  {journal} {\bibinfo  {journal}
  {Sov. Phys. Uspekhi}\ }\textbf {\bibinfo {volume} {8}},\ \bibinfo {pages}
  {743} (\bibinfo {year} {1966})}\BibitemShut {NoStop}%
\bibitem [{ {Aue}\ \emph {et~al.}(1976) {Aue},
   {Bartholdi},\ and\  {Ernst}}]{Aue76}%
  \BibitemOpen
  \bibfield  {author} {\bibinfo {author} { {W.~P.}\ 
  {Aue}}, \bibinfo {author} { {E.}~ {Bartholdi}}, \
  and\ \bibinfo {author} { {R.~R.}\  {Ernst}},\ }\href
  {\doibase http://dx.doi.org/10.1063/1.432450} {\bibfield  {journal} {\bibinfo
   {journal} {J. Chem. Phys.}\ }\textbf {\bibinfo {volume} {64}},\ \bibinfo
  {pages} {2229} (\bibinfo {year} {1976})}\BibitemShut {NoStop}%
\bibitem [{ {Kainosho}\ \emph {et~al.}(2006)
  {Kainosho},  {Torizawa},  {Iwashita}, 
  {Terauchi},  {Mei~Ono},\ and\ 
  {Guntert}}]{Kainosho06}%
  \BibitemOpen
  \bibfield  {author} {\bibinfo {author} { {M.}~
  {Kainosho}}, \bibinfo {author} { {T.}~ {Torizawa}},
  \bibinfo {author} { {Y.}~ {Iwashita}}, \bibinfo
  {author} { {T.}~ {Terauchi}}, \bibinfo {author}
  { {A.}~ {Mei~Ono}}, \ and\ \bibinfo {author}
  { {P.}~ {Guntert}},\ }\href
  {http://dx.doi.org/10.1038/nature04525} {\bibfield  {journal} {\bibinfo
  {journal} {Nature}\ }\textbf {\bibinfo {volume} {440}},\ \bibinfo {pages}
  {52} (\bibinfo {year} {2006})}\BibitemShut {NoStop}%
\bibitem [{ {Kessler}\ \emph {et~al.}(2013)
  {Kessler},  {Komar},  {Bishof}, 
  {Jiang},  {Sorensen},  {Ye},\ and\ 
  {Lukin}}]{Kessler13x}%
  \BibitemOpen
  \bibfield  {author} {\bibinfo {author} { {E.~M.}\ 
  {Kessler}}, \bibinfo {author} { {P.}~ {Komar}},
  \bibinfo {author} { {M.}~ {Bishof}}, \bibinfo
  {author} { {L.}~ {Jiang}}, \bibinfo {author}
  { {A.~S.}\  {Sorensen}}, \bibinfo {author}
  { {J.}~ {Ye}}, \ and\ \bibinfo {author}
  { {M.~D.}\  {Lukin}},\ }\href@noop {} {\bibfield
  {journal} {\bibinfo  {journal} {ArXiv e-prints},\ \bibinfo {pages}
  \Eprint {http://arxiv.org/abs/1310.6043} {arXiv:1310.6043}} (\bibinfo {year} {2013})}\BibitemShut {NoStop}%
\bibitem [{ {Trusheim}\ \emph {et~al.}(2013)
  {Trusheim},  {Li},  {Laraoui}, 
  {Chen},  {Gaathon},  {Bakhru}, 
  {Schroder},  {Meriles},\ and\ 
  {Englund}}]{Trusheim13}%
  \BibitemOpen
  \bibfield  {author} {\bibinfo {author} { {M.~E.}\ 
  {Trusheim}}, \bibinfo {author} { {L.}~ {Li}},
  \bibinfo {author} { {A.}~ {Laraoui}}, \bibinfo
  {author} { {E.~H.}\  {Chen}}, \bibinfo {author}
  { {O.}~ {Gaathon}}, \bibinfo {author} {
  {H.}~ {Bakhru}}, \bibinfo {author} {
  {T.}~ {Schroder}}, \bibinfo {author} { {C.~A.}\
   {Meriles}}, \ and\ \bibinfo {author} {
  {D.}~ {Englund}},\ }\href {\doibase 10.1021/nl402799u} {\bibfield
   {journal} {\bibinfo  {journal} {Nano Letters}\ }\textbf {\bibinfo {volume}
  {1}},\ \bibinfo {pages} {32} (\bibinfo {year} {2013})}\BibitemShut {NoStop}%
\bibitem [{ {Neumann}\ \emph {et~al.}(2010)
  {Neumann},  {Beck},  {Steiner}, 
  {Rempp},  {Fedder},  {Hemmer}, 
  {Wrachtrup},\ and\  {Jelezko}}]{Neumann10b}%
  \BibitemOpen
  \bibfield  {author} {\bibinfo {author} { {P.}~
  {Neumann}}, \bibinfo {author} { {J.}~ {Beck}},
  \bibinfo {author} { {M.}~ {Steiner}}, \bibinfo
  {author} { {F.}~ {Rempp}}, \bibinfo {author}
  { {H.}~ {Fedder}}, \bibinfo {author} {
  {P.~R.}\  {Hemmer}}, \bibinfo {author} {
  {J.}~ {Wrachtrup}}, \ and\ \bibinfo {author} {
  {F.}~ {Jelezko}},\ }\href {\doibase 10.1126/science.1189075}
  {\bibfield  {journal} {\bibinfo  {journal} {Science}\ }\textbf {\bibinfo
  {volume} {5991}},\ \bibinfo {pages} {542} (\bibinfo {year}
  {2010})}\BibitemShut {NoStop}%
\bibitem [{ {Ajoy}\ and\ 
  {Cappellaro}(2012)}]{Ajoy12g}%
  \BibitemOpen
  \bibfield  {author} {\bibinfo {author} { {A.}~
  {Ajoy}}\ and\ \bibinfo {author} { {P.}~
  {Cappellaro}},\ }\href {\doibase 10.1103/PhysRevA.86.062104} {\bibfield
  {journal} {\bibinfo  {journal} {Phys. Rev. A}\ }\textbf {\bibinfo {volume}
  {86}},\ \bibinfo {pages} {062104} (\bibinfo {year} {2012})}\BibitemShut
  {NoStop}%
\bibitem [{ {Ohno}\ \emph {et~al.}(2012) {Ohno},
   {{Joseph Heremans}},  {Bassett}, 
  {Myers},  {Toyli},  {{Bleszynski Jayich}},
   {Palmstrom},\ and\  {Awschalom}}]{Ohno12}%
  \BibitemOpen
  \bibfield  {author} {\bibinfo {author} { {K.}~
  {Ohno}}, \bibinfo {author} { {F.}~ {{Joseph
  Heremans}}}, \bibinfo {author} { {L.~C.}\ 
  {Bassett}}, \bibinfo {author} { {B.~A.}\  {Myers}},
  \bibinfo {author} { {D.~M.}\  {Toyli}}, \bibinfo
  {author} { {A.~C.}\  {{Bleszynski Jayich}}},
  \bibinfo {author} { {C.~J.}\  {Palmstrom}}, \ and\
  \bibinfo {author} { {D.~D.}\  {Awschalom}},\ }\href
  {\doibase 10.1063/1.4748280} {\bibfield  {journal} {\bibinfo  {journal}
  {Appl. Phys. Lett.}\ }\textbf {\bibinfo {volume} {101}},\ \bibinfo {pages}
  {082413} (\bibinfo {year} {2012})}\BibitemShut {NoStop}%
\bibitem [{ {Osterkamp}\ \emph {et~al.}(2013)
  {Osterkamp},  {Scharpf},  {Pezzagna}, 
  {Meijer},  {Diemant},  {Jurgen~Behm}, 
  {Naydenov},\ and\  {Jelezko}}]{Osterkamp13}%
  \BibitemOpen
  \bibfield  {author} {\bibinfo {author} { {C.}~
  {Osterkamp}}, \bibinfo {author} { {J.}~ {Scharpf}},
  \bibinfo {author} { {S.}~ {Pezzagna}}, \bibinfo
  {author} { {J.}~ {Meijer}}, \bibinfo {author}
  { {T.}~ {Diemant}}, \bibinfo {author} {
  {R.}~ {Jurgen~Behm}}, \bibinfo {author} {
  {B.}~ {Naydenov}}, \ and\ \bibinfo {author} {
  {F.}~ {Jelezko}},\ }\href {\doibase
  http://dx.doi.org/10.1063/1.4829875} {\bibfield  {journal} {\bibinfo
  {journal} {Appl. Phys. Lett.}\ }\textbf {\bibinfo {volume} {103}},\ \bibinfo
  {eid} {193118} (\bibinfo {year} {2013})}\BibitemShut {NoStop}%
\bibitem [{ {Naydenov}\ \emph {et~al.}(2010)
  {Naydenov},  {Richter},  {Beck}, 
  {Steiner},  {Neumann},  {Balasubramanian},
   {Achard},  {Jelezko},  {Wrachtrup},\
  and\  {Kalish}}]{Naydenov10}%
  \BibitemOpen
  \bibfield  {author} {\bibinfo {author} { {B.}~
  {Naydenov}}, \bibinfo {author} { {V.}~ {Richter}},
  \bibinfo {author} { {J.}~ {Beck}}, \bibinfo {author}
  { {M.}~ {Steiner}}, \bibinfo {author} {
  {P.}~ {Neumann}}, \bibinfo {author} {
  {G.}~ {Balasubramanian}}, \bibinfo {author} {
  {J.}~ {Achard}}, \bibinfo {author} {
  {F.}~ {Jelezko}}, \bibinfo {author} {
  {J.}~ {Wrachtrup}}, \ and\ \bibinfo {author} {
  {R.}~ {Kalish}},\ }\href {\doibase 10.1063/1.3409221} {\bibfield
  {journal} {\bibinfo  {journal} {App. Phys. Lett.}\ }\textbf {\bibinfo
  {volume} {96}},\ \bibinfo {eid} {163108} (\bibinfo {year}
  {2010})}\BibitemShut {NoStop}%
\bibitem [{ {Naydenov}\ \emph {et~al.}(2011)
  {Naydenov},  {Dolde},  {Hall}, 
  {Shin},  {Fedder},  {Hollenberg}, 
  {Jelezko},\ and\  {Wrachtrup}}]{Naydenov11}%
  \BibitemOpen
  \bibfield  {author} {\bibinfo {author} { {B.}~
  {Naydenov}}, \bibinfo {author} { {F.}~ {Dolde}},
  \bibinfo {author} { {L.~T.}\  {Hall}}, \bibinfo
  {author} { {C.}~ {Shin}}, \bibinfo {author}
  { {H.}~ {Fedder}}, \bibinfo {author} {
  {L.~C.~L.}\  {Hollenberg}}, \bibinfo {author} {
  {F.}~ {Jelezko}}, \ and\ \bibinfo {author} {
  {J.}~ {Wrachtrup}},\ }\href {\doibase 10.1103/PhysRevB.83.081201}
  {\bibfield  {journal} {\bibinfo  {journal} {Phys. Rev. B}\ }\textbf {\bibinfo
  {volume} {83}},\ \bibinfo {pages} {081201} (\bibinfo {year}
  {2011})}\BibitemShut {NoStop}%
\bibitem [{ {de~Lange}\ \emph {et~al.}(2010)
  {de~Lange},  {Wang},  {Rist\`{e}}, 
  {Dobrovitski},\ and\  {Hanson}}]{DeLange10}%
  \BibitemOpen
  \bibfield  {author} {\bibinfo {author} { {G.}~
  {de~Lange}}, \bibinfo {author} { {Z.~H.}\  {Wang}},
  \bibinfo {author} { {D.}~ {Rist\`{e}}}, \bibinfo
  {author} { {V.~V.}\  {Dobrovitski}}, \ and\ \bibinfo
  {author} { {R.}~ {Hanson}},\ }\href {\doibase
  10.1126/science.1192739} {\bibfield  {journal} {\bibinfo  {journal}
  {Science}\ }\textbf {\bibinfo {volume} {330}},\ \bibinfo {pages} {60}
  (\bibinfo {year} {2010})}\BibitemShut {NoStop}%
\bibitem [{ {{Romach}}\ \emph {et~al.}(2014)
  {{Romach}},  {{Muller}},  {{Unden}}, 
  {{Rogers}},  {{Isoda}},  {{Itoh}}, 
  {{Markham}},  {{Stacey}},  {{Meijer}},
   {{Pezzagna}},  {{Naydenov}}, 
  {{McGuinness}},  {{Bar-Gill}},\ and\ 
  {{Jelezko}}}]{Romach14}%
  \BibitemOpen
  \bibfield  {author} {\bibinfo {author} { {Y.}~
  {{Romach}}}, \bibinfo {author} { {C.}~ {{Muller}}},
  \bibinfo {author} { {T.}~ {{Unden}}}, \bibinfo
  {author} { {L.~J.}\  {{Rogers}}}, \bibinfo {author}
  { {T.}~ {{Isoda}}}, \bibinfo {author} {
  {K.~M.}\  {{Itoh}}}, \bibinfo {author} {
  {M.}~ {{Markham}}}, \bibinfo {author} {
  {A.}~ {{Stacey}}}, \bibinfo {author} {
  {J.}~ {{Meijer}}}, \bibinfo {author} {
  {S.}~ {{Pezzagna}}}, \bibinfo {author} {
  {B.}~ {{Naydenov}}}, \bibinfo {author} { {L.~P.}\
   {{McGuinness}}}, \bibinfo {author} {
  {N.}~ {{Bar-Gill}}}, \ and\ \bibinfo {author} {
  {F.}~ {{Jelezko}}},\ }\href@noop {} {\bibfield  {journal}
  {\bibinfo  {journal} {ArXiv e-prints}\ } (\bibinfo {year} {2014})},\ \Eprint
  {http://arxiv.org/abs/1404.3879} {arXiv:1404.3879} \BibitemShut
  {NoStop}%
\bibitem [{ {Suzuki}(1976)}]{Suzuki76}%
  \BibitemOpen
  \bibfield  {author} {\bibinfo {author} { {M.}~
  {Suzuki}},\ }\href {http://dx.doi.org/10.1007/BF01609348} {\bibfield
  {journal} {\bibinfo  {journal} {Comm in Mat.h Phys.}\
  }\textbf {\bibinfo {volume} {51}}},\ \bibinfo {pages} {183} (\bibinfo {year}
  {1976})\BibitemShut {NoStop}%
\bibitem [{ {Haeberlen}\ and\ 
  {Waugh}(1968)}]{Haeberlen68}%
  \BibitemOpen
  \bibfield  {author} {\bibinfo {author} { {U.}~
  {Haeberlen}}\ and\ \bibinfo {author} { {J.}~
  {Waugh}},\ }\href {\doibase 10.1103/PhysRev.175.453} {\bibfield  {journal}
  {\bibinfo  {journal} {Phys. Rev.}\ }\textbf {\bibinfo {volume} {175}},\
  \bibinfo {pages} {453} (\bibinfo {year} {1968})}\BibitemShut {NoStop}%
\end{thebibliography}
%

\end{article}  
\end{document}